\documentclass[epj]{svjour}

\usepackage{graphics}
\usepackage{url}

\usepackage{amsmath,amsthm,amssymb,bm,braket} 
\usepackage{xcolor}
\newcommand{\oprt}[1]{\ensuremath{\hat{\mathcal{#1}}}}
\newcommand{\abs}[1]{\ensuremath{\left| #1 \right|}}
\def \beq{\begin{equation}}
\def \eeq{\end{equation}}
\def \beqa{\begin{eqnarray}}
\def \eeqa{\end{eqnarray}}
\def \Schr{Schr\"odinger }

\newcommand{\be}{\begin{equation}}  
\newcommand{\ee}{\end{equation}}
\newcommand{\bc}{\begin{center}}
\newcommand{\ec}{\end{center}}

\begin{document}
\title{An overview of the scientific contribution of Andrea Vitturi to Nuclear Physics}
\subtitle{being an account of the recent TNP19 meeting held in Padova.}
\author{L. Fortunato\inst{1,2} \and  
	C.E. Alonso \inst{3,4} \and 
	J.M. Arias \inst{3,4} \and 
	J. Casal \inst{1,2} \and 
	K. Hagino \inst{5,6,7} \and 
	J.A. Lay \inst{3,4} \and 
	E.G. Lanza \inst{8,9} \and 
	S.M. Lenzi \inst{1,2} \and 
	J. Lubian \inst{10} \and 
	T. Oishi \inst{11} \and 
	F. P\'erez-Bernal \inst{12,4}  
}                     
\institute{
Dip. Fisica e Astronomia ``G. Galilei", Univ. Padova, Padova, Italy \and  
I.N.F.N. Sez. di Padova, via Marzolo, 8, I-35131 Padova, Italy \and
Dep. de F{\'\i}sica At\'omica, Molecular y Nuclear, Univ. Sevilla, Spain \and
Instituto Interuniversitario Carlos I de F\'\i sica Te\'orica y Computacional (iC1), Apdo.~1065, E-41080 Sevilla, Spain \and
Department of Physics, Tohoku University, Sendai 980-8578, Japan \and 
Research Center for Electron Photon Science, Tohoku University, 1-2-1 Mikamine, Sendai 982-0826, Japan \and
Department of Physics, Kyoto University, Kyoto 606-8502, Japan \and 
I.N.F.N. - Sezione di Catania, via S. Sofia 64, I-95123 Catania, Italy  \and
Dipartimento di Fisica e di Astronomia ``Ettore Majorana", Univ. Catania, Italy \and
Instituto de F\'isica, Universidade Federal Fluminense, 24210-340, Niter\'oi, Rio de Janeiro, Brazil \and 
Department of Physics, Faculty of Science, University of Zagreb, Bijeni\v{c}ka c. 32, 10000 Zagreb, Croatia \and 
Depto. de Ciencias Integradas y Centro de Estudios Avanzados en F\'\i sica, Matem\'aticas y Computaci\'on (CEAFMC), Universidad de Huelva, 21071 Huelva, Spain.
 }
\date{Received: date / Revised version: date}
%
\abstract{
We give an account of the main achievements of the scientific career of Andrea Vitturi so far,  that have recently been discussed during the workshop ``Theoretical Nuclear Physics in Padova" on the occasion of his retirement from full professor at the University of Padova. He has oftentimes been the driving force behind numerous contributions to nuclear structure and nuclear reactions that are here reviewed: giant resonances, pairing correlations, collective modes, algebraic models, inelastic excitations, electromagnetic response, break-up and transfer reactions, coupled-channel formalism, clustering, subbarrier fusion processes, etc.
Among these topics several inspirational works and ideas can be found that we would like to highlight.
}
\PACS{
      {PACS-key}{discribing text of that key}   \and
      {PACS-key}{discribing text of that key}
     } 
%
\maketitle

This manuscript serves two scopes. On one hand, it is an account of some of the topics discussed in the recent workshop ``Theoretical Nuclear Physics in Padova: a meeting in honor of Prof. Andrea Vitturi"\cite{TNP19} that was held in Padova on May 20 and 21 2019 on occasion of his retirement. On the other, the organizers and participants felt the duty to summarize in this review paper his extensive scientific production, beyond what was discussed at the meeting, because it spans a broad range of topics in low energy nuclear physics, trying to highlight several fruitful ideas and to shed new light on some less known papers that most certainly deserve further recognition.

\section{Alma mater and vita in brief}
Padova is the seat of one of the most ancient universities in the world, founded in 1222 by a group of scholars leaving from Bologna\footnote{According to Wikipedia, where several historical references are collected, Padova is the 5th oldest university in continuous operation, after Bologna, Oxford, Salamanca and Cambridge. The counting raises to 8th if one includes institutions that have been closed.}.
Physics at the University of Padova also has a very ancient tradition, in fact Galileo Galilei, the tuscanian proposer of the scientific method, inventor of the telescope, discoverer of several laws of Nature and one of the fathers of modern physics (together with Kepler, Newton and many others), had been a professor in Padova for 18 years.
More modernly, Nuclear Physics has represented a major asset of cutting edge research at the University of Padova, supported in this by the establishment in 1968, very close to Padova, of the Legnaro National Laboratories, one of four Labs of the Italian National Institute of Nuclear Physics (I.N.F.N.)\footnote{The others being located in Catania, Frascati and Gran Sasso.}.
\begin{figure}
\bc
\resizebox{.4\columnwidth}{!}{\includegraphics{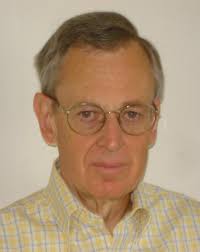}}~
\resizebox{.5\columnwidth}{!}{\includegraphics{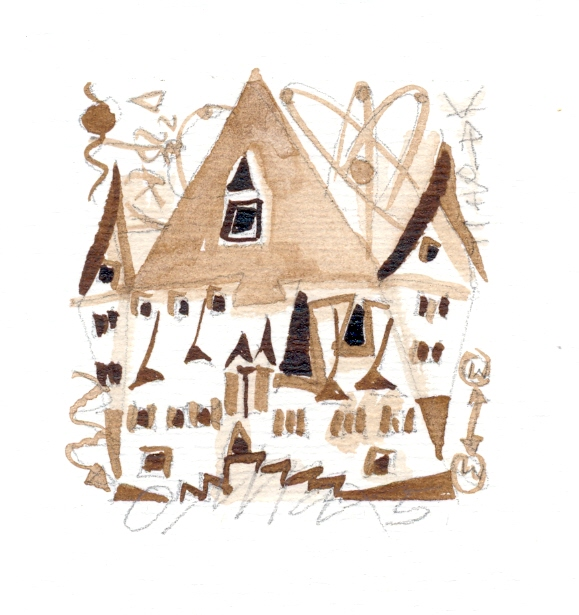}}
\ec
\caption{Prof. Andrea Vitturi and logo of the conference organized in Fiera di Primiero in 2007, featuring the Palazzo delle Miniere, i.e. the location of the event (Artist: Max Art).}
\label{fig:1}      
\end{figure}

Andrea Vitturi was born in Farra di Soligo, province of Treviso, in 1949 and studied physics at the University of Padova, culminating in a degree thesis with honours in 1972 under the supervision of Claudio Villi with Franco Zardi as co-supervisor. The latter had been a researcher in theoretical nuclear physics for the local branch of the I.N.F.N. and collaborated closely with Andrea Vitturi in the first part of his scientific career, especially on the formal theory of nuclear resonances and on Glauber theory. The first publication on the latter subject came in 1977 on the journal ``Lettere al Nuovo Cimento" \cite{1977Vi06} with title ``Analysis of 1.37 GeV alpha-12C Elastic Scattering in Terms of a Modified Eikonal Multiple-Scattering Theory". 

He then spent a few years, from 1977 to 1979, as a Research Associate at the Niels Bohr Institute in Copenhagen, where the local school of nuclear theory was very prominent, working in particular with Aage Winther, Ricardo Broglia and several younger colleagues of about the same age, who then became long-standing collaborators and friends. He specialized in reaction theory, heavy-ion physics, coupled-channel calculations, inelastic excitations and transfer reactions on one side and in nuclear structure, in particular in the newly formulated theory that connects pairing interaction and collective models. 

He then returned to Italy as a research staff member at the University of Padova from 1981 to 1987 and he became associate professor at the University of Trento (Italy) in 1988, where he studied collective modes of motion, giant resonances and algebraic models. He moved to Padova in 1991 and became full professor of nuclear physics in 2001. He has had several students, visitors and collaborators and, among other tasks, served as director of the Ph.D. School in Physics and as head of the theory group for a few years. He has also been Chairman of the Program Advisory Committee at the LNL in Legnaro. While in Padova he focused in processes involving nuclei far from the stability valley on one side and in quantum shape phase transitions on the other side and the intersections of these topics with pairing and transfer reactions.

He has been appointed as visiting scientist or visiting professor in several prestigious institutions, the University of Seville, Oak Ridge National Laboratories, the University Paris Sud at Orsay, the University of Tsukuba, the UFF in Rio de Janeiro and the ECT* in Trento, just to name a few. He has also been the driving force behind a very successful series of biannual meetings, called ``Selected topics in Nuclear and Atomic physics", the first editions of which were organized in Varenna, a beautiful spot on the Como lake, and then, from 1999, moved to Fiera di Primiero, a wonderful little hamlet among the Dolomite mountains in northern Italy.

\subsection{Paper's organization}

Vitturi's scientific production has been copious and varied, more than 200 items according to citation databases \cite{NSR,Scopus}, ranging on such a large number of topics that the inclusion of several coauthors has been necessary both to share the load and to render justice to it without leaving behind any important topic.

We have tried to summarize the papers where new ideas or new solutions to given problems are proposed. Sometimes the developments in these papers have proven crucial to their respective sub-fields and sometimes they only partially have got the recognition that they would deserve. Despite the fact that alphabetically he is very often at the end of the list of authors, Vitturi's humble attitude often prevented him from taking full credit of the fact that he has frequently been the driving force beyond many developments. Therefore this paper aims also at shedding new light on older works and set them in the right perspective. Notwithstanding the traditional division into nuclear structure and nuclear reactions will be used here for the sake of organizing the material in a logical way, the philosophy of Andrea Vitturi and many of his collaborators has always been that these two aspects are strongly intertwined and they must be viewed as two faces of the same coin: we cannot investigate nuclear structure properties without subjecting the nuclear system to some kind of process that ultimately necessitates of a proper description of its dynamics. This has been a constant {\it Streben} in the scientific works of Andrea Vitturi. He has made a lifelong effort to warn his peers about the fact that, when trying to extract structure properties, such as transition probabilities or matrix elements, from reaction cross-sections, one might fall into silly blunders. The pitfall, very often, is the too optimistic separation of reaction cross-section into factors that separate the reaction part from the structure part.
Under certain conditions this separation can be achieved, while in general, care should be taken by assessing the proper conditions that allow, for example, the extraction of multipole strength functions from inelastic cross-sections or two-particle transition strengths from transfer cross-sections.

Without respecting the chronological order, we will discuss studies on quadrupole collectivity and quantum shape phase transitions in Sect. \ref{Quad} and \ref{QPT}.
Then, we will discuss Giant Resonances in Sect. \ref{GR} and Giant Pairing Vibrations in Sect. \ref{GPV}
Two-neutron pairing correlations will be discussed in Sect. \ref{TNPC} and then we will move to one-dimensional structure and reaction models (Sect. \ref{1D}) and treat the correlations of two fermions in the continuum in Sect. \ref{2Fcont}, \ref{PC} and  \ref{D2nTR}. Sect. \ref{2N} will be devoted to various aspects of structure and dynamics of two-neutron transfer reactions. 
Sect. \ref{eikonal} will summarize studies on the eikonal approximation in treating direct reactions at intermediate energies.
The relevance of low-lying continuum in photo-excitation processes and in break-up reactions at low-energy will be discussed in Sect. \ref{Cont-Clus} as well as studies related to clustering. Subbarrier fusion reactions and fusion of clusterized systems at astrophysical energies will be treated in Sect. \ref{FUS} and \ref{Astro} respectively.

\section{Quadrupole collective model} \label{Quad}
This is a very complex topic and we do not attempt a systematic summary of all its ramifications, we will limit ourselves to highlight the contributions that came from the groups in Padova and in Seville, where Andrea Vitturi gave a very important contribution. 
The quadrupole degree of freedom is one of the most important collective modes of motion of nuclei, it consists of vibrations and rotations of an ellipsoidally deformed shape around an equilibrium configuration (that might be spherical or already deformed). 

\subsection{Solutions of the Bohr hamiltonian} \label{BH}
\begin{figure*}[tb]
	\begin{center}
		\resizebox{1.3\columnwidth}{!}{ \includegraphics{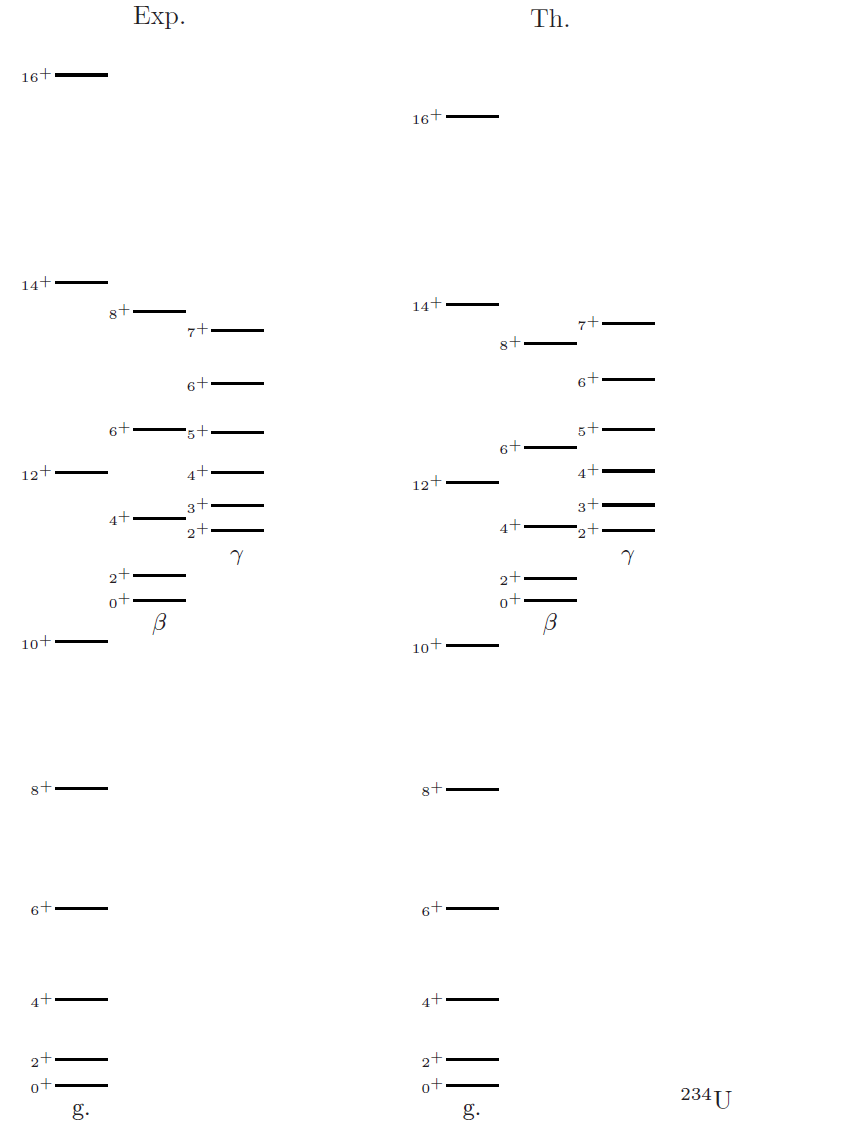}}
	\end{center}
	\caption{
		Comparison between the calculated and experimental spectra of $^{234}$U, ground-state band and one-phonon $\beta-$ and $\gamma-$vibrations. The theoretical predictions are obtained with the choice $B = 393.7415$ and $c = 41.5629$ in Eq. (\ref{eq:Kra}) and (\ref{eq:HO}). From Ref. \protect\cite{2004FOxx}.}
	\label{fig:234U}       
\end{figure*}

After the inception of the critical point symmetries E(5) and X(5) by Francesco Iachello \cite{2000IA,2001IA} the field of applications of the Bohr hamiltonian to study collective quadrupole spectra of even-even nuclei got a new boost. The Bohr-Mottelson model describes the most important mode of motion of nuclei at low energy, that consists in rotations and vibrations of the quadrupole deformed nuclear surface.
Andrea Vitturi has been very active in this field and stimulated new developments from his students.
This led to two exact solutions of the Bohr hamiltonian with a gamma-independent Coulomb and a Kratzer potential that can be condensed in the following expression:
\be
V(\beta) = \frac{A}{\beta}+  \frac{B}{\beta^2}
\label{eq:Kra}
\ee
where the Coulomb potential can be recovered by setting $B=0$. This potential \cite{2003FO08}, once plugged into the five-dimensional Schr\"odinger equation for the Bohr hamiltonian leads to a separable exact solution. The nice feature is that the eigenvalues and eigenvectors can be written exactly in a very compact fashion in terms of special functions that can be used as bases for diagonalization of more complex problems. These two exact solutions are part of a small club of exact models, together with Bohr-and-Mottelson's harmonic oscillator potential, Elliott's Davidson potential and Iachello's square-well E(5) potentials. This field of studies then exploded, much beyond the purest intentions of the authors of these solutions, generating a plethora of models and solutions schemes.
In a subsequent paper \cite{2004FOxx}, that was drawing the approximate separation of variables from the X(5) model of Iachello, the Bohr hamiltonian was solved with Coulomb and Kratzer potentials in $\beta$ plus an harmonic oscillator in the $\gamma$ variable, namely 
\be
V(\gamma)= c \gamma^2/2 \;.
\label{eq:HO}
\ee
As an example of application, we give in Fig. \ref{fig:234U} the two-parameter fit of the Kratzer potential plus gamma-soft harmonic oscillator to the low-lying energy spectrum of $^{234}$U. The ground state band as well as the gamma- and beta-bands are excellently fitted with such a simple model, and allow for reliable predictions. 
We are confining ourselves here to the works where Andrea Vitturi has given a direct contribution. The history of exact and approximate solutions of the Bohr hamiltonian is partly summarized in several review papers (cfr. \cite{2005FO18}, \cite{buganu}, \cite{Casten} and \cite{Cejnar}) and the innumerable applications to spectroscopy are still flourishing today. We do not attempt to cover the whole issue and we refer the reader to the specialized literature, and clearly this means leaving aside important developments in this field, developed by several groups around the world. 

\begin{figure*}
	\bc
	\resizebox{1.4\columnwidth}{!}{%
		\includegraphics{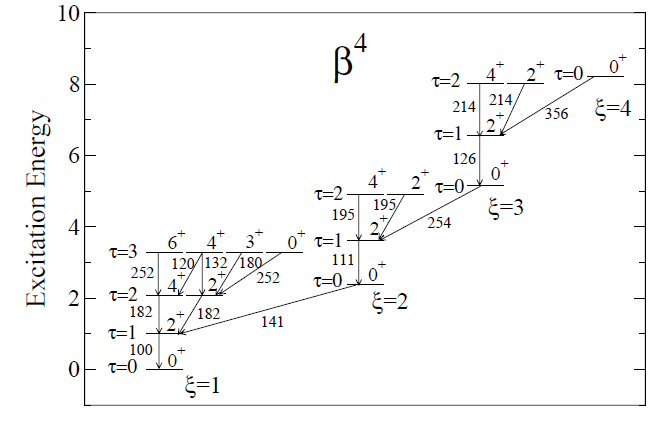}
	}\ec
	\caption{Energy spectrum of a quartic potential, $\beta^4$. Quantum numbers $\xi$ and $\tau$ are given, together with spin-parity $J^\pi$. The numbers close to arrows are $B(E2)$ values relative to the $2_1^+\rightarrow 0_1^+$ transition that is normalized to 100. Figure taken from Ref. \protect\cite{2003AR24}.}
	\label{fig:beta4}       
\end{figure*}

One case of particular interest \cite{2003AR24} is the numerical solution of the quartic potential $V(\beta) = A \beta^4$ that is what one expects from IBM and coherent states for the critical point along a phase transition connecting the spherical U(5) regime and the axially deformed SU(3) limit. The authors of \cite{2003AR24} showed that the ratio $R_{4/2}$ is equal to 2.09 in this case, quite different from the prediction of the E(5) model, that is 2.20. In Fig. \ref{fig:beta4} the lowest part of the energy spectrum for such a potential is shown. States are labelled by quantum numbers $\xi, \tau$ and $L$ associated with the $\gamma-$independent symmetry. This spectrum is qualitatively quite similar to the E(5) case, but important differences are seen not only in the energies, but also in the transition rates.

The advantage of choosing the square well in the E(5) model is clearly that it is amenable to exact solution, but the quartic solution is possibly closer to the true behaviour of a second order phase transition from spherical to deformed phases (Landau-type potential). A more general version of this potential containing quadratic, quartic as well as cubic terms, has been applied also to the X(5) candidate nucleus $^{138}$Gd \cite{2012Pr00} and in other studies. Other extensions with higher powers in $\beta$ have been worked out in Ref. \cite{Bonatsos,Bonatsos2}.

\section{Quantum Phase Transitions in IBM and IBFM} \label{QPT}
This topic runs in parallel with the previous one, the main difference being that the quadrupole degree of freedom is explored within algebraic models such as the Interacting Boson Model for even-even systems  \cite{IBMbook} and the Interacting Boson-Fermion Model for odd-even systems \cite{IBFMbook}. These approaches based on symmetry and Lie algebras allow for in-depth connection of the algebraic interacting boson models with the collective models. In addition, during recent years the Padova-Sevilla collaboration has worked together in the study of structural phase shape transitions in IBM and IBFM in relation to the so-called critical point symmetries \cite{2000IA,2001IA,X5*03}. The reader that is unfamiliar with the basics of this topic might find an elementary introduction in Ref. \cite{Euro}, where the jargon and main ideas are explained.

\subsection{Connection of the IBM and IBFM to the collective model by using the intrinsic frame formalism} 
The traditional approach to connect the algebraic IBM and IBFM models with the more traditional geometrical collective models has been through the intrinsic state formalism \cite{GK80}. This formalism allows to extract a potential energy surface from a given Hamiltonian as depending on shape variables. It is a variational method in which one proposes an intrinsic state for the system that depends on a few parameters that have to be determined by minimizing the expectation value of the corresponding Hamiltonian in the intrinsic state. Within the IBM, the intrinsic state for the ground state band for an even-even nucleus is written as a boson condensate of the form
\begin{equation}
|\Phi_{gs}(\beta,\gamma)\rangle =\frac{1}{\sqrt{N_B!}} ~[\Gamma^{\dag}_{gs}(\beta,\gamma)]^{N_B}|0\rangle ,
\label{p_int}
\end{equation}
where $N_B$ is the boson number, $|0\rangle$ is the boson vacuum and the ground state boson creation
operator, $\Gamma^{\dag}_{gs}(\beta,\gamma)$, is
\begin{equation}
\Gamma^{\dag}_{gs}(\beta,\gamma)=\frac{1}{\sqrt{1+\beta^2}} [s^\dag+\beta \cos\gamma d^{\dag}_{0} \\
+ \frac{\beta}{\sqrt{2}} \sin\gamma(d^{\dag}_{2}+d^{\dag}_{-2})].
\label{b_int}
\end{equation}
The potential energy surface for the ground state is obtained by calculating the expectation value of the appropriate boson Hamiltonian, $H_B$, in the intrinsic state
(\ref{p_int})
\begin{equation}
E_{gs}(\beta,\gamma)=\langle\Phi_{gs}(\beta,\gamma)|H_B|\Phi_{gs}(\beta,\gamma) \rangle .
\label{E_int}
\end{equation}
The variational parameters $\beta$ and $\gamma$ play a similar role to the ones of the intrinsic collective shape variables in the Bohr Hamiltonian.

Intrinsic frame states for the mixed boson-fermion systems can be constructed by coupling the odd single-particle states to the intrinsic states of the even core. The lowest states of the odd nucleus are expected to originate from the above mentioned coupling to the intrinsic ground-state $|\Phi_{gs}(\beta,\gamma)\rangle$. To obtain them, one should first construct the coupled states
\begin{equation}
|\Psi_{jK}(\beta,\gamma)\rangle= |\Phi_{gs}(\beta,\gamma)\rangle \otimes |jK\rangle
\label{P_int}
\end{equation}
and then diagonalize the total boson-fermion Hamiltonian in this basis, giving a set of energy eigenvalues $E_n$($\beta,\gamma$), where $n$ is an index to count solutions in the odd-even system. In the present case, for an angular momentum $j$, the possible magnetic components are denoted by $K$. Therefore there are $2j+1$ different states that are restricted due to the symmetry  $ K \leftrightarrow -K$.

Using this formalism with the appropriate extensions, the Padova-Sevilla collaboration has studied many different aspects of the even-even and odd-even systems and the relationships of these algebraic models with the traditional geometric models. This has been a long-standing and very fruitful collaboration and we just mention some results obtained in these studies, referring the reader to the specialized literature for more details:

\begin{enumerate}
	\item even-even $O(6) \leftrightarrow \gamma-$unstable: quadrupole moments, E2 transitions and transition densities in the O(6) limit of the interacting boson model were studied within a formalism based on the intrinsic frame \cite{QE2gamma88,TDgamma89}
	
	\item odd-even systems: within the framework of an intrinsic frame formalism both axially symmetric and triaxial situations were discussed. The role of Bose-Fermi symmetries and their interpretation in terms of shape variables were studied. Also the structure of $\beta$ and $\gamma$ excitations in odd-even systems was investigated \cite{Franco92}.
	
	\item octupole deformations and coupling dipole modes to quadrupole deformations: an intrinsic frame analysis of octupole deformed nuclei in the SU(3) limit of the extended Interacting Boson Model was presented. Excited bands associated with nuclei exhibiting permanent octupole deformation were studied, as well as the behavior of in-band and intra-band transitions. The coupling of an odd particle to an even core with these characteristics was also discussed \cite{Octupole95}. In similar spirit, the coupling of the high-lying dipole mode to the low-lying quadrupole modes for the case of deformed $\gamma$-unstable nuclei was studied \cite{Dipole01}. 
	
	\item cranking: a self-consistent cranking formalism based on the interacting boson model intrinsic wave function was developed. Explicit formulae were obtained for the energies and moments of inertia of the ground-state rotational band as a power expansion in the rotational frequency \cite{cranking96}.
	
	\item two-phonon states: a general study of excitations up to two-phonon states was carried out using the intrinsic-state formalism of the Interacting Boson Model (IBM). Spectra and transitions for the different dynamical symmetries were analyzed and the correspondence with states in the laboratory frame was established \cite{twophonon98}.
	
	\item mixed symmetry states and $\alpha-$transfer: within the neutron-proton interacting boson model the population of mixed-symmetry states via alpha transfer processes was studied.  Closed expressions were deduced in the case of the limiting U$_{\pi + \nu }(5)$ and SU$_{\pi + \nu }(3)$ symmetries.  It was found that the population of the lowest mixed symmetry $2^+$ state, vanishing along the $N_{\pi}=N_{\nu}$ line, depends on the number of active bosons and it is normally smaller than the lowest fully symmetric $2^+$ state \cite{alphatransfer08}.
	
	\item triaxiality: an extension of the Interacting Boson Model that includes the cubic $(\hat Q \times \hat Q \times \hat Q)^{(0)}$ term was proposed. The potential energy surface for the cubic quad\-ru\-pole interaction was explicitly calculated within the coherent state formalism using the complete ($\chi-$de\-pen\-dent) expression for the quad\-ru\-pole operator. The Q-cubic term was found to depend on the asymmetry deformation parameter $\gamma$ in a combination of $\cos{(3\gamma)}$ and $\cos^2{(3\gamma)}$ terms, thereby allowing for minima in the triaxial region ($0^o < \gamma_0 < 60^o$) of the parameter space \cite{QQQ11}. A similar form of cubic terms inducing triaxiality, based on $d$ operators, were introduced in the IBM by Heyde and collaborators \cite{1984He}.
\end{enumerate}

\subsection{Shape phase transitions in the IBM and the IBFM} 
With the introduction of the idea of critical point symmetries by Iachello in early 2000 \cite{2000IA,2001IA}, a new field of research concerning critical point symmetries started and this reopened the way to studies on structural phase transitions in Nuclear Physics. The works of the collaboration including Andrea Vitturi in this field can be summarized as follows:

\begin{enumerate}
	\item even-even $U(5) \rightarrow O(6)$ critical point vs. $E(5)$: the differences between the E(5) critical point introduced by Iachello within the collective model  \cite{2000IA} and the equivalent critical point in the IBM \cite{E5beta403} have been critically discussed. In addition, the validity of the intrinsic state formalism for transitional IBM regions \cite{Inci09} was tested. 
	
	\item one-particle spectroscopic intensities in $\gamma-$unstable systems: the evolution of one-particle spectroscopic factors as a possible signature of shape phase transitions was investigated. The study was done by describing the odd systems in terms of the Interacting Boson Fermion Model. The particular case of an odd j=3/2 particle coupled to an even-even boson core\footnote{It is well-known (Bayman-Silverberger) that the coupling of a $\gamma$-unstable core with a $j=3/2$ orbits leads to a supersymmetric situation, i.e. it can be described with a Lie superalgebra.} that undergoes a phase transition from spherical U(5) to $\gamma$-unstable O(6) situation was considered \cite{Spect06}.
	
	\item two-particle transfer as a tool to detect shape phase transitions: the evolution of the transfer spectroscopic intensities as a possible signature of shape-phase transitions was investigated. The study was carried out considering chains of even-even nuclei displaying changes in shape, such as from sphericity to axial-symmetric deformed or from sphericity to deformed $\gamma-$unstable nuclei. The evolution of the structure of these nuclei was described in terms of the interacting boson model. Simple formulae were given using the intrinsic-frame formalism \cite{Ruben07}.
	
	\item odd-nuclei and critical point symmetries in IBFM: several different aspects connected with shape phase transitions in nuclei and the possible occurrence of dynamical symmetries at the critical points including multiorbit symmetries have been investigated. The phase transition in odd nuclei when the underlying even-even core nuclei experiment transitions between two dynamical symmetries of the types 
  $U(5) \rightarrow O(6)$,  $U(5) \rightarrow SU(3)$ and  $SU(3) \rightarrow O(6) \rightarrow \overline{SU(3)}$
	were studied. The odd particle was assumed to be moving in different single particle orbitals \cite{QPToddgamma05,OddMultiorbit07,U5SU3odd09,Mahmut10,Mahmut11,Mahmut15}. At the critical point in the phase transition spherical$-\gamma$-unstable for odd-even nuclei, an analytic multi-orbit solution to the corresponding Bohr Hamiltonian, called E(5/12), was worked out \cite{E51207}.
	
\end{enumerate}

 \begin{figure*}	\bc
 	\resizebox{1.\textwidth}{!}{%
 		\includegraphics{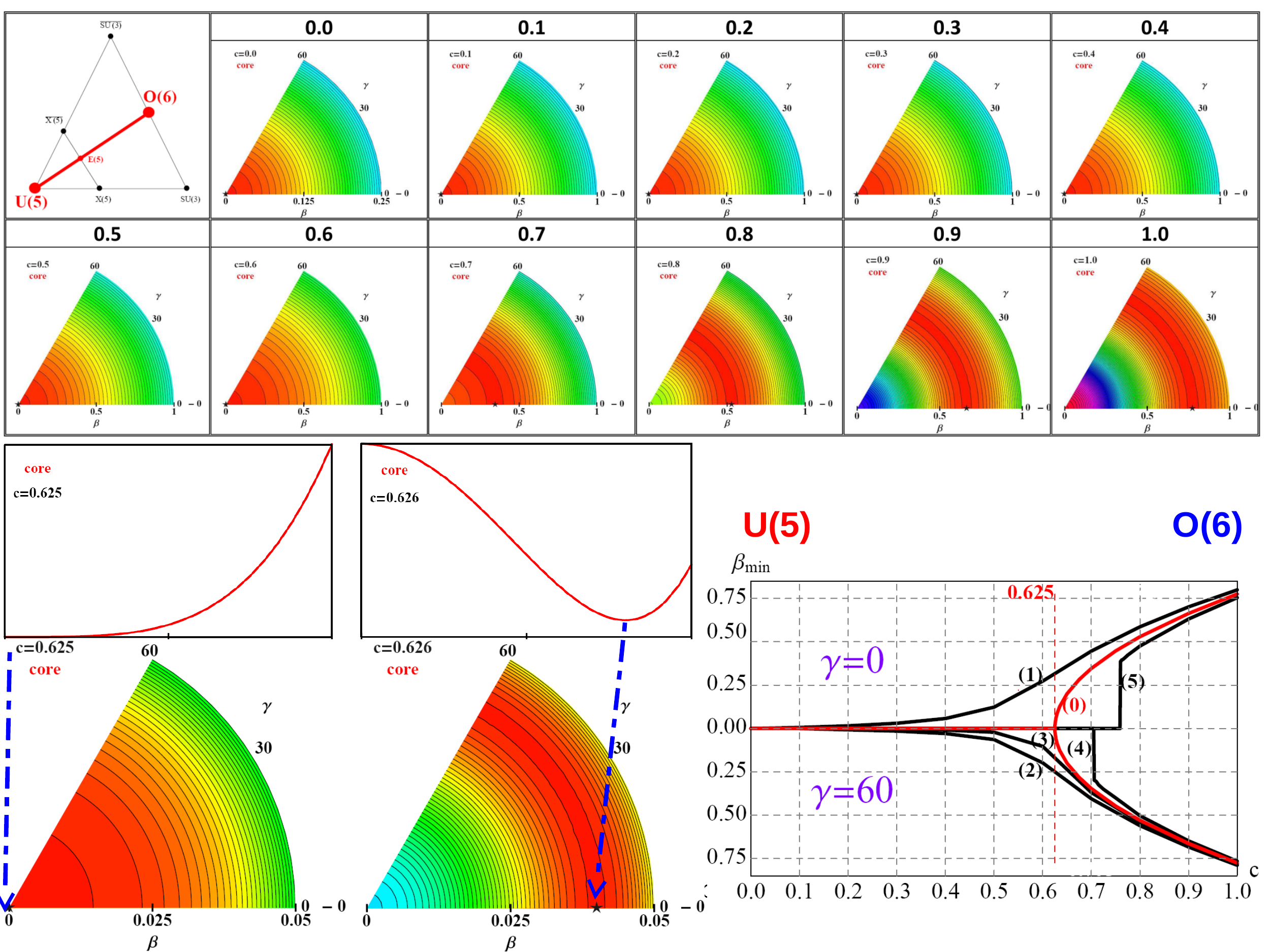}
 	}\ec
 	\caption{Example of a $j=9/2$ particle coupled to a even-even core undergoing the transition from spherical to deformed $\gamma-$unstable shapes. For more details see text.}
 	\label{fig:1pepe}       
 \end{figure*}
 
 As an example of the obtained results for odd-even systems in the vicinity of the critical point in the even nuclei, we discuss briefly the case of the transition from vibrational behaviour to the gamma-unstable regime (characterized within the collective Bohr hamiltonian by the E(5) critical point symmetry).  Along this line, a simple parameterization for the hamiltonian of the even-even system is used, namely
 \begin{equation}
 H_B= (1-c) ~\hat n_d - \frac{c}{4N} \hat Q_B \cdot \hat Q_B ~,
 \label{HQQ}
 \end{equation}
 that is often called the consistent Q hamiltonian \cite{book_cas}, where
 \begin{eqnarray}
 \hat n_d & = & \sum_\mu d^\dag _\mu d_\mu ~~, \\ 
 \hat Q_B  & = & (s^\dag \times \tilde d + d^\dag \times \tilde s+\chi~ d^\dag \times \tilde d )^{(2)} ~~,
 \label{QB}
 \end{eqnarray}
 and $N$ is the total number of bosons. With the choice $\chi=0$ this hamiltonian produces a transition between spherical and $\gamma-$unstable phases, the critical point being obtained for $c=N/(2(N-1))$. The spectrum at the critical point is qualitatively similar to the one obtained with the E(5) model in solution of the Bohr hamiltonian. Similarly, one can describe the corresponding phase transitions in the neighbouring odd nuclei.  This is formally obtained in the Interacting Boson Fermion Model (IBFM) by a boson-fermion hamiltonian of the type
 \begin{equation}
 H~=~H_{B}+H_{F}+V_{BF}~,
 \label{HBF2}
 \end{equation}
 where
 \begin{equation}
 H_F+V_{BF}~=~\sum{_j}~\epsilon_j a^\dag_j \cdot a_j~-~
 2 \frac{c}{4N} \hat Q_B \cdot \hat q_F~.
 \label{VBF1}
 \end{equation}
 $\epsilon_j$ are the single particle energies, and $\hat q_F$ is the fermion quadrupole operator. With the choice $\chi=0$ for the boson quadrupole operator one can describe, by varying the mixing parameter $c$, the transition from sphericity to deformed gamma-instability. In Fig. \ref{fig:1pepe} the odd particle is assumed to be moving in the single particle orbital $j=9/2$. In the left panel of the upper part, the path followed when changing $c$ is plotted. Note that we have written $E(5)$ in the diagram as a reference, but this is not the IBM critical point as shown in Ref. \cite{E5beta403}. All the other panels in the upper part are the energy surfaces for the even-even core for different $c-$values. The critical point is at $c=0.625$ as shown in the lower left panels. The lower right panel gives the behavior of the single particle states: $(1) \equiv K=\pm 9/2$, $(2) \equiv K=\pm 7/2$,  $(3) \equiv K=\pm 5/2$,  $(4) \equiv K=\pm 3/2$,  $(5) \equiv K=\pm 1/2$. The line $(0)$ corresponds to the behavior of the ground state energy for the even-even core.
 The various components of the single-particle odd state are driven to the prolate or oblate side at different values of the control parameter, depending on the value of $K$ and the preferential alignment of angular momenta.

\section{Giant Resonances} \label{GR}
Giant resonances are a universal feature of the spectrum of nuclei, arising in photo-absorption, inelastic scattering, charge-exchange and other processes. They are collective mode of oscillation of the nucleus as a whole, excited either by the electromagnetic field or by the strong force, and they can be seen in a fluid dynamics picture, but of a different kind with respect to surface oscillations described in the previous sections. Microscopically, they are understood as a coherent superposition of particle-hole excitations in a shell-model picture of the nucleus. Andrea Vitturi has specialized in the field of giant resonances while at Copenhagen, describing the optimal conditions for exciting this mode \cite{1979BR0x,1980BR0x} and continued in Trento and Padova, where he investigated also aspects related to excitation of double and multiphonon giant resonances \cite{1987CA0x,2003DA0x} and to new phenomena arising in neutron-rich nuclei (to be discussed in the next section).

\subsection{Giant and pigmy resonances}
During recent years considerable interest has been focused on the study of
the effect of neutron excess on the collective properties of neutron-rich
nuclei. Due to the presence of the so-called neutron skin one would expect an
enhanced mixing of the isoscalar and isovector modes\footnote{These are modes of motion in which the proton and neutron fluids oscillate in phase or out-of-phase, respectively.}.

To investigate this point, microscopic calculations
for Oxygen and Calcium isotopes,
based on spherical Hartree-Fock model with Skyrme SGII
interaction have been performed\cite{cat97a,cat97b}.  
The collective excitations of these nuclei were determined in the Random Phase Approximation (RPA),
using the full residual interaction. The strength distributions for the dipole mode were calculated for multipole operators of the form 
\be
O^{(EM)}_{1M}= {eN\over A}
\sum_{p=1}^Z r_p Y_{1M}(\hat r_p) - {eZ\over A} \sum_{n=1}^N r_n Y_{1M}(\hat r_n) \>
\ee
for the electromagnetic dipole and 
\be
O^{(IS)}_{1M}= \sum_{i=1}^A (r_1^3 - {5 \over 3}<r^2> r_i)Y_{1M}(\hat r_i) \>
\ee 
for the $3\hbar \omega$ dipole mode of isoscalar nature. The analytic form of the  operator in the second equation above avoids the spurious center of mass motion.

These calculations have shown that as soon as the neutron number
increases, some strength appears at low energies in the dipole
strength distribution, well below the dipole giant resonance. 
\begin{figure*}\bc
	\resizebox{1.4\columnwidth}{!}{%
		\includegraphics{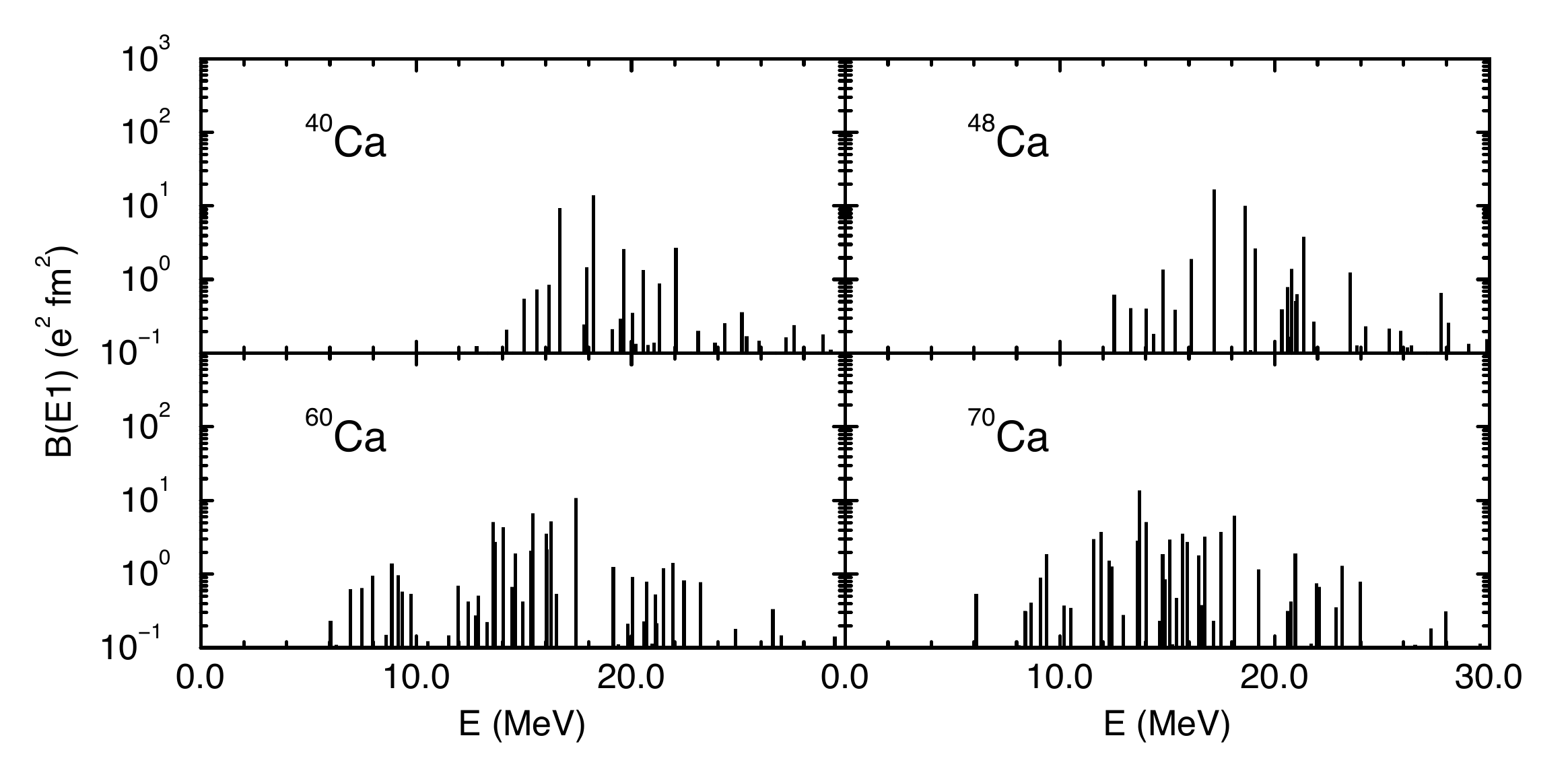}	}\ec
	\caption{Isovector dipole response obtained in HF+RPA for a sequence of a Calcium isotopes.}
	\label{lanza1}      
\end{figure*} 

\begin{figure*}\bc
	\resizebox{1.4\columnwidth}{!}{%
		\includegraphics{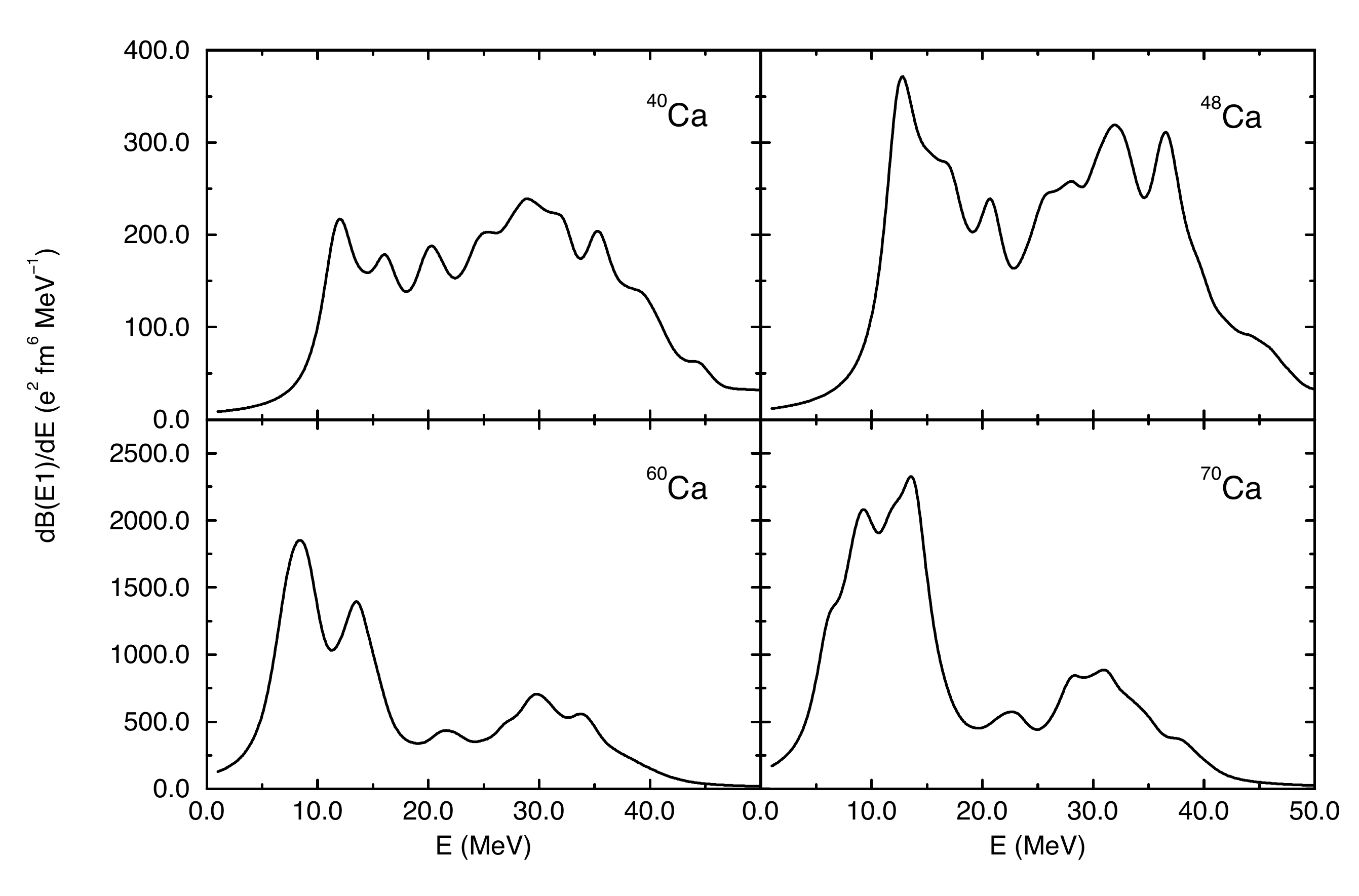}
	}\ec
	\caption{Averaged isoscalar dipole response obtained in HF+RPA for a sequence of a Calcium isotopes.}
	\label{lanza2}       
\end{figure*}

This can be seen in fig.~\ref{lanza1}, where the isovector response obtained in a HF + RPA calculation with the SGII Skyrme interaction are plotted.
The strength appearing at low energy, carrying few per cent of the isovector EWSR, is present in
many nuclei far from the stability line and it has been associated to the possible existence of another mode of new nature: the Pygmy Dipole Resonance (PDR)\cite{paa07,sav13,bra15,bra19}.
In fig.~\ref{lanza2}  the isoscalar response is shown for the same Calcium isotopes of fig.~\ref{lanza1}, in this case the strength has been averaged with a Lorentzian function with a width of 2 MeV. The isoscalar response shows a behaviour similar to the isovector one, even though a peak is present also for nuclei with N=Z. The increasing number of neutron moves the isoscalar strength to lower energies until the peak at lower energy becomes stronger than the one corresponding to the isoscalar Giant Dipole Resonance (ISGDR). 

\begin{figure}
	\resizebox{.8\columnwidth}{!}{%
		\includegraphics{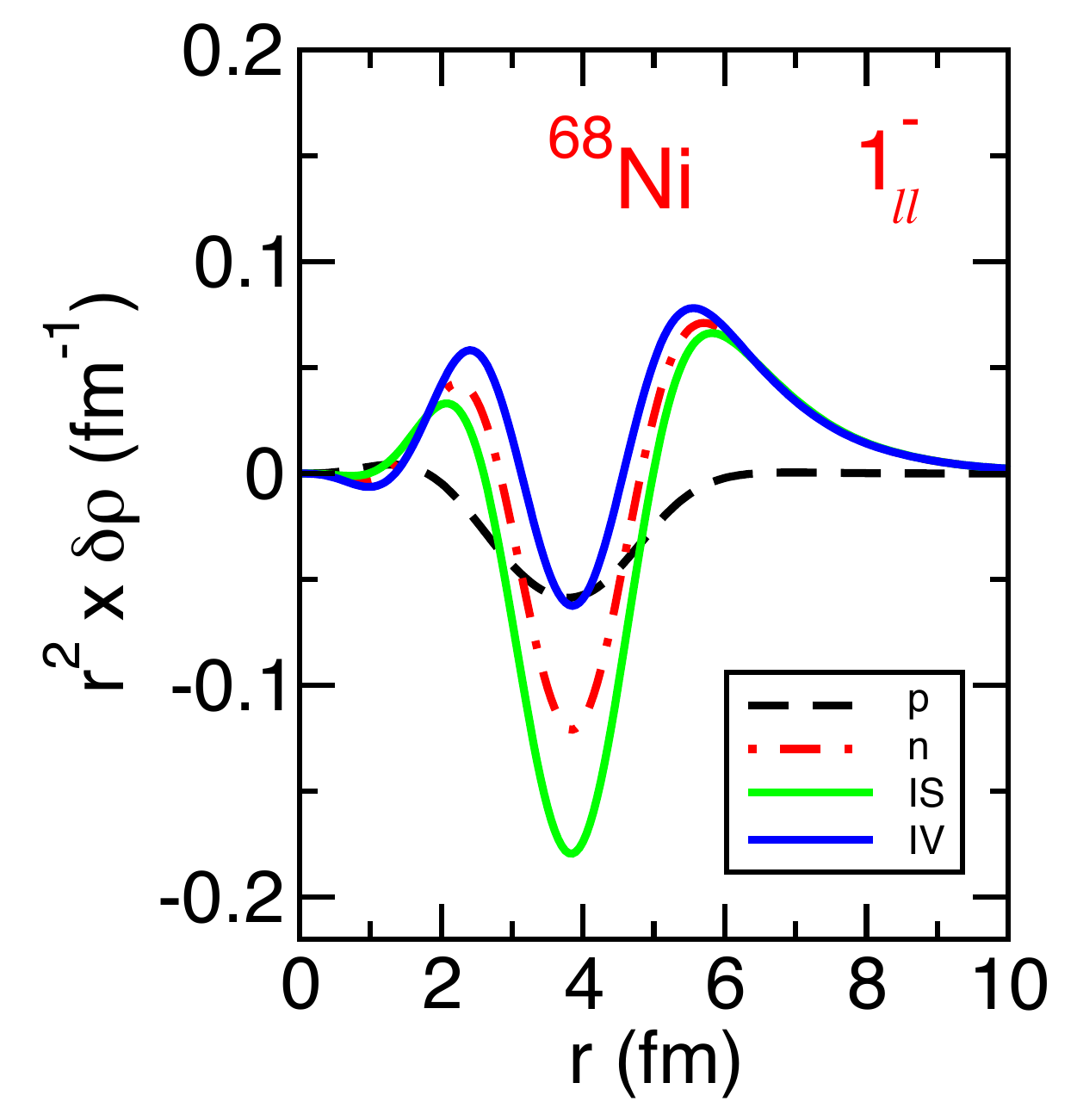}
	}
	\caption{Proton (dashed black) and neutron (dot-dashed red) radial transition densities for the pygmy dipole state at E=10.41 MeV of the $^{68}$Ni isotope. The microscopic transition densities have been obtained by a HF plus RPA calculation with an SGII interaction. The isovector (solid black) and isoscalar (solid green) transition densities are also shown.}
	\label{lanza3}       
\end{figure}

The nature of the different states and the effect of the neutron excess
are best evidenced by looking at the transition densities of these states.
In the case of the dipole, the proton and neutron components of the
transition densities are in phase inside the nucleus while at
the nuclear surface the neutrons give practically the only
contribution, and therefore isoscalar and isovector transition
densities are of comparable radial dependence and magnitude, see fig.~\ref{lanza3}.
One therefore should expect that these states, due to their mixing of isoscalar and isovector
character, will respond in an equivalent way to isoscalar and
isovector heavy-ion probes that are only sensitive to the surface\cite{lan11,2011VI0x}.

This has been shown by many experiments using the ($\alpha, \alpha' \gamma$) reactions as well as other isoscalar probes like $^{17}$O\cite{bra15}. Experiments performed on the same nuclei with the ($\gamma, \gamma'$) reaction reveal an interesting property of the low-lying dipole states: the states belonging to the lower part of the PDR energy region are excited by both isoscalar and isovector probes, while the states lying in the high energy part are excited only by the electromagnetic probe. This clear separation of the PDR energy range in two regions is  generally referred to as the splitting of the PDR. This result is achieved by comparing the reduced electromagnetic transition probability B(E1) with the inelastic cross section measured with isoscalar probes. However, while the Coulomb inelastic cross section and the B(E1) are proportional, the relationship between the isoscalar response and the inelastic excitation cross section due to an isoscalar probe it is not so evident. In Ref. \cite{lan14}, semiclassical cross section calculations have been performed that reproduce the global feature of the reduction of the experimental cross section at higher excitation energy compared to the isovector channel, confirming the structural splitting of the low-lying E1 strength.

The isospin mixing of the low-lying dipole states has strong consequences in the excitation process especially when an isoscalar probe is used. One of the most important ingredients entering in the calculation of the inelastic cross section is the radial form factor which contains the structure effects and depends on the various model employed for the description of the inner structure of the nuclei.  In ref.~\cite{lan15} it has been shown that the form factor constructed via a double folding procedure are very sensitive to the shape of the transition densities employed, especially in the region around the nuclear surface which is the region explored in such nuclear reactions. Therefore, the choice of a transition density which describes all the details of the state taken into consideration is of paramount importance.

\section{Giant Pairing Vibrations} \label{GPV}
The Giant Pairing Vibration (GPV) is an elusive collective mode of motion of nuclei, that is connected with the fields of giant resonances (from which it draws the first part of the name) and of  pairing interactions in nuclei. It was predicted long ago \cite{1977BR0x} in a collective model created by analogy with the Bohr-Mottelson quadrupole hamiltonian, but for the monopole degree of freedom. The B\'es hamiltonian \cite{1970Be0x}, as it is called, describes the quantized pairing oscillations and rotations in an abstract two-dimensional gauge space. In a polar parameterization this space can be described by two variables, the radial variable $\alpha$ that plays the role of pairing collective variable, controlling the extent of "deformation" in this space and a gauge angle $\phi$. The B\'es hamiltonian is a fruitful model that allows to explore two limits, depending on the shape of a collective potential $V(\alpha)$, that might range from harmonic oscillator with a minimum in $\alpha=0$, describing "normal nuclei" or "spherical in gauge space", around which pairing vibrations might occur, to potentials with a minimum for $\alpha > 0$, describing "superfluid nuclei" or "deformed in gauge space" around which one expects a  pairing rotational spectrum to occur. The spectra we are talking about include the monopole states around the $0+$ ground state of an even-even nucleus, including a chain of neighbouring even-even nuclei that differ by an even amount of particles from the initial nucleus.
The occurrence of a sort of harmonic scheme around $^{208}$Pb was pointed out by Bohr and Mottelson \cite{BM book}. Pairing rotations are instead associated with superfluid beahviour, and in bewteen these extremes one has also a critical point solution of the B\'es hamiltonian that gives rise to interesting speculations \cite{2006CL01}.

Instead of opting for collective approaches, one might also look at microscopic models, in which the addition or removal mode are described in terms of particle-particle or hole-hole operators tensorially coupled to zero spin, $[a^\dag a^\dag]^{0}_0$ and $[aa]^{0}_0$ that create or destroy a monopole pair respectively. These can be plugged into a particle-particle Random Phase Approximation framework \cite{RS book,1970Be0x,1971Be,1971VA}, where it is possible to calculate pair transfer transition densities that are a way to characterize the amount of pairing correlations in each state (pp-RPA). Performing a BCS  transformation on top of RPA calculations allows for an extension to superfluid systems. In this case addition and removal modes cannot be calculated separately, but the $X$ and $Y$ coefficients should be sought in the solution of the dispersion equation.

The history of the search for this mode of motion is nicely collected in two recent papers \cite{2016BO25,2019AS0x}, where the connections with collective models and with microscopic models that make use of second quantization are explored and where the links of these studies with two-neutron transfer reactions are summarized. 
An early review of these topics, that has the merit of giving detailed explanations that are hard to find elsewhere, was edited by Andrea Vitturi, Carlos Dasso and joined many collaborators in the proceedings of a meeting promoted by the Italian Physical Society ({\it Societ\`a Italiana di Fisica}, SIF)  held at the Villa Monastero in Varenna (Como lake district, Italy) in 1987 \cite{1987DV-book}.
\begin{figure*}[tb]
	\begin{center}
		\resizebox{1.5\columnwidth}{!}{ \includegraphics{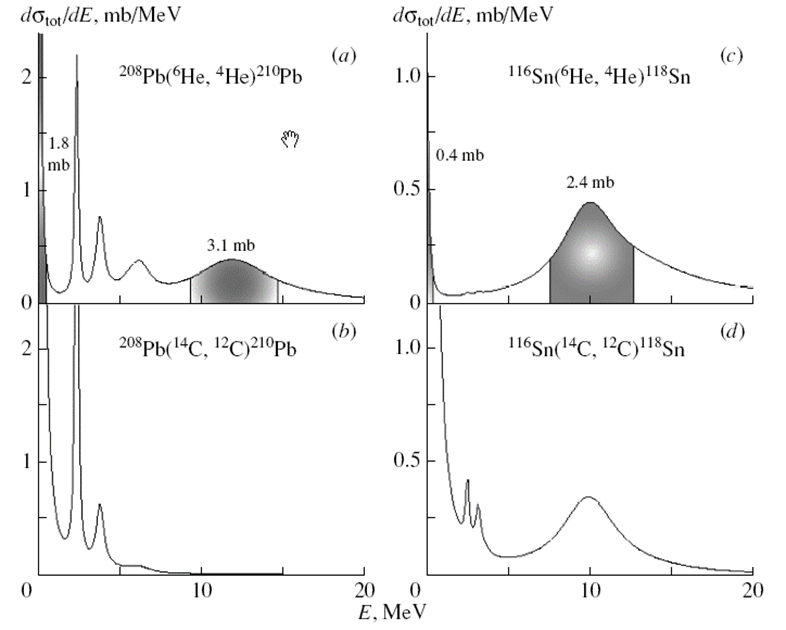}}
	\end{center}
	\caption{
		Comparison of calculated DWBA differential cross-sections to ground and excited states of $^{210}$Pb and $^{118}$Sn (calculated in a ppRPA approach, addition mode) in two-neutron transfer reactions with conventional beams (lower panels) and weakly-bound nuclei (upper panels). The excitation of the GPV mode is favoured in the latter case, Adapted from Ref. \protect\cite{2002FO08}.}
	\label{fig:gpvtheo}       
\end{figure*}
The contribution of Andrea Vitturi to this field has been very important, especially in collaboration with Hugo Sofia, showing that the number non conserving nature of this mode implies that it can be observed in reactions that allow for a transfer or an exchange back and forth of nucleons and that the properties of these reactions depend not only on the structure properties of the mode itself, but also, and heavily, on the dynamical variables associated with bombarding energy, choice of target-projectile combinations \cite{2015DA0x} and especially on the reaction Q-value. This last point, in particular, has been the subject of several investigations aimed at finding an explanation of the reason why several experimental campaigns in the '70's and more recently  failed to observe the GPV in heavy nuclei with $(t,p)$ reactions. In Ref. \cite{2002FO08}, a particle-particle RPA approach was used for tin and lead nuclei, with the aim of calculating the pairing strengths $\beta_p$ of ground- and excited $0^+$ states, in which the GPV is identified as the collective state (large $\beta_p$) within the group of excited level with 2 quanta of excitation around an energy of about $2\hbar\omega$. These model structure calculation, however imperfect, have been merged with reactions calculations, employing either a standard projectile, like $^{14}$C, or an exotic weakly-bound nucleus, namely $^6$He. The intuition of Andrea Vitturi and Hugo Sofia has been that the weakly-bound nature of this light Borromean nucleus, implies an optimum Q-value that will favour the population of monopole modes at around the GPV energy, depressing instead the population of the ground state, a fact that is impossible to obtain with conventional beams of well bound projectiles or with triton beams (See Fig. \ref{fig:gpvtheo}). This study and subsequent additions based on the same idea \cite{2003FO12} led to a reconsideration of the optimal experimental conditions that must be sought to effectively prove the existence of the GPV. Finally after more than 40 years of efforts, the first serious hints of having achieved this have been obtained by the MAGNEX group at the {\it Laboratori Nazionali del Sud} in Catania \cite{2015CA01}. 
Measurements of differential cross section in two-neutron transfer reactions with $^{18}$O impinging on $^{14}$C and $^{15}$C have shown clear signatures of an excited monopole mode ($L=0$) with high collective character at about the energy predicted by simple formulas. The excitation spectrum for $^{14}$C and the angular distribution of the proposed GPV state are shown in Fig. \ref{fig:gpvexp}.
\begin{figure*}[tb]
	\begin{center}
		\resizebox{.9\textwidth}{!}{ \includegraphics{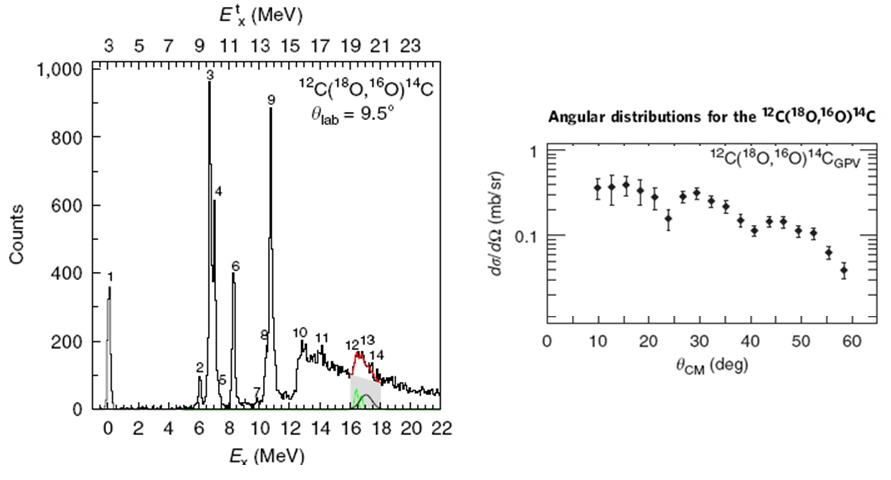}}
	\end{center}
	\caption{Energy spectrum for the reaction $^{12}$C($^{18}$O,$^{16}$O)$^{14}$C showing a peak at about 16.9 MeV of excitation energy. The angular distribution for this peak shows an oscillatory pattern that indicates monopole character, according to the analysis discussed in Fig. 3 and in the Supplementary material of Ref. \cite{2015CA01}. Figure remastered from Ref. \protect\cite{2015CA01}.}
	\label{fig:gpvexp}       
\end{figure*}
This important study suggests that the expected similarity between p-h states in giant resonances and the corresponding p-p or h-h collective GPV states, far from being just a formal analogy, approximately holds true, despite the GPV being far less accessible to experimental investigation.
In evaluating the transfer probability and quantal corrections, the contribution of Andrea Vitturi has been very important \cite{2001VO20}.
The elusive appearance of the GPV mode in reactions with light nuclei was further investigated in Ref. \cite{2016Las}, where the role of low-angular momentum orbitals and the Q-value mismatch are emphasized.

\section{One-dimensional models for structure and reactions} \label{1D}
In the spirit of understanding all the ingredients of physical models, before attempting application to real systems, Andrea Vitturi suggested to work on the line of research of one-dimensional models for structure and reactions of weakly-bound and borromean nuclei. The previous experience of one of the authors (F.P.B.) in dealing with a pseudostate approach to continuum discretization with a Transformed Harmonic Oscillator basis was first applied to 1D problems \cite{PBernal2001} and later to 3D problems \cite{Moro2006}. 
The 1D model was instrumental for grasping the subtle features that arise from the coupling of a weakly-bound quantum system with the continuum. The model was a
one-dimensional three-body model, inspired to Ref. \cite{Bertsch1991},
that consisted on two valence unbound neutrons interacting with a Woods-Saxon potential
core with completely filled bound levels. In this approach the two
nucleons, with coordinates \(x_1\) and \(x_2\), are bound because they
interact with each other via a density-dependent contact residual interaction
\(V_{int}(x_1,x_2)\) of the type

\begin{equation}
V_{int}(x_1,x_2) = -V_{RI} \left[\frac{\rho[(x_1+x_2)/2]}{\rho_0}\right]^p \delta(x_1-x_2),
\label{rint}
\end{equation}
\noindent where  \(V_{RI}\), \(p\), and \(\rho_0\) are free parameters while
\(\rho(x)\) is the matter density of the \(N_b\) occupied bound states of the core, namely  \(\rho(x) = \sum_{i=0}^{N_b-1}\psi^*_i(x)\psi_i(x)\).

The work on the dynamics of this model, with the extremely valuable
participation of Kouichi Hagino and Hiro Sagawa, gave as a result the
publication of an article where the effect of dineutron correlation on
breakup processes were studied by subjecting them to the influence of a
time-dependent one-body external field \cite{Hagino2010}.
This work also influenced the subsequent work of Oishi and collaborators, summarized in section \ref{2Fcont}.

The work in collaboration with Andrea Vitturi continued, resulting in several
contributions to congresses and meetings and later it was extended to Laura Moschini, one of Vitturi's graduate students, who in her PhD thesis and the subsequent publications, studied the aforementioned model in a systematic and exhaustive way, comparing the
outcome of different methods of discretizing the continuum with
pseudostates and the pros and cons of each of them
\cite{Moschini2016}. Direct reactions, such as elastic and inelastic excitations  \cite{1985LA0x}, transfer and breakup can be studied in this model \cite{2015Vi00,2018Mo00}, that contains most of the features encountered in real systems and allows for a number of interesting speculations about the reaction mechanisms. For example, the transfer of two particles is clearly enhanced in collision processes and the ability of several models to give accurate results can be tested against exact solutions in the 1D formalism.

The solution of the time-dependent Schr{\"o}dinger equation can be compared with coupled-channel formalism with and without the inclusion of the continuum, thus allowing for a quantitative evaluation of the effectiveness of each approach. A review of these studies is given in Ref. \cite{2017Vi00}.

\section{Correlations of two fermions in the continuum} \label{2Fcont}
Pairing correlation is one of the most curious phenomena 
of multi-fermion systems, including atomic nuclei. 
This correlation plays an essential role not only 
in static systems, but also in dynamical phenomena. 
One example of these phenomena is two-nucleon radioactive 
emission \cite{2009GRXX,2012PFXX}, 
where a pair of proton or neutron is emitted 
by the quantum-tunneling effect. 
Key ingredients of this process include 
(i) the pairing correlation, 
(ii) Pauli principle, and 
(iii) time-dependent dynamics in the continuum region. 
For a proper description of the two-nucleon emission, 
one should simultaneously consider these ingredients 
in the system of interest. 
For this purpose, time-dependent (TD) multi-particle models provide 
a suitable solution, see Ref. \cite{2018OI03} and references therein.

In this section, we introduce the basic idea and 
formalism of TD model. 
We also review the results of TD calculation applied to the 
two-fermion (2F) tunneling in one-dimension \cite{2018OI03}, 
where the physical properties of two-nucleon 
radioactivity can be characterized.

\subsection{Time-dependent calculations}
Quantum-mechanical dynamics is an essential concept to 
understand various phenomena of the nuclear and subatomic physics. 
Those are generally described by the 
time-dependent \Schr equation, 
\begin{equation}
i\hbar \frac{\partial}{\partial t} \ket{\psi(t)} = \oprt{H} \ket{\psi(t)}, 
\end{equation}
where $\oprt{H}$ is the total Hamiltonian corresponding to 
the system of interest. 
For simplicity,  the Hamiltonian $\oprt{H}$ is assumed to be
static in the following. 
Also, in this subsection, the specific form of $\oprt{H}$ is not specified, in order to keep the formalism very general.

Employing the eigenstates of $\oprt{H}$ as basis, namely
$\oprt{H} \ket{E} = E \ket{E}$,  allows to write an arbitrary state $\ket{\psi_0}$ as an expansion 
\beq
\ket{\psi_0} = \int \mu (E) \ket{E} dE, \label{eq:init_c}
\eeq
where $\left\{ \mu(E) \right\} \in \mathbb{C}$ 
are the expansion coefficients. 
Thus, $\abs{\mu(E)}^2$ indeed corresponds to the energy 
spectrum. 
Assuming $\ket{\psi_0}$ as the initial state, 
the time-evolution of this state can be found by 
\beq
\ket{\psi (t)} 
=e^{-it\oprt{H}/\hbar} \ket{\psi_0 } 
=\int \mu (E) e^{-itE/\hbar} \ket{E} dE. \label{eq:psi_t}
\eeq
The survival coefficient, $\beta(t)$, is determined 
as the overlap between the initial and the present states. 
That is, 
\beqa
\beta(t) &\equiv & \Braket{\psi_0 | \psi(t)} \nonumber \\
&=& \int dE' \mu (E') \int dE \mu (E) \Braket{E' | e^{-itE/\hbar} | E} \nonumber \\
&=& \int dE \abs{\mu (E)}^2 e^{-itE/\hbar}, \label{eq:Krylov}
\eeqa
where the notation $\Braket{E' \mid E}=\delta(E'-E)$ was used.  
From Eq. (\ref{eq:Krylov}), one can read that 
the survival coefficient is given by the Fourier transformation of 
the energy spectrum. 
This is nothing but the ``Krylov-Fock theorem'' \cite{1989KuXX}. 
From $\beta(t)$, the survival probability $P_{\rm surv}(t)$ is also determined as 
\beq
P_{\rm surv}(t) = \abs{\beta(t)}^2, \label{eq:TDP}
\eeq
which physically corresponds to the decaying rule of this meta-stable state.

In many cases of nuclear radioactive processes, the decaying rule 
can be well approximated by an exponential function: 
\beq
P_{\rm surv}(t) \cong e^{-t/\tau}, \label{eq:EXP_PS}
\eeq
where $\tau$ is the lifetime of the meta-stable state. 
From Eqs. (\ref{eq:Krylov}) and (\ref{eq:TDP}), 
the corresponding spectrum to this exponential decay 
can be found in terms of the Breit-Wigner (BW) profile. 
That is, 
\beq
\abs{\mu (E)}^2 = \frac{1}{\pi} \frac{(\Gamma_0 /2)}{(E-E_0)^2 + (\Gamma_0 /2)^2}, \label{eq:BW1}
\eeq
or equivalently, 
\beq
\ket{\psi_0} = \int  \sqrt{\frac{\Gamma_0}{2\pi}} 
\frac{e^{ia(E)} }{(E_0 - i\Gamma_0/2) - E} \ket{E}   dE, \label{eq:BW2}
\eeq
where $\left\{ e^{ia(E)} \right\}$ with $a(E) \in \mathbb{R}$ are 
arbitrary phase-factors. 
The width $\Gamma_0$ is related to the lifetime, $\tau=\hbar /\Gamma_0$. 
The central energy $E_0$, on the other hand, corresponds to the 
experimental Q value of the process of interest.

\subsection{Application to two-fermion tunneling}
\begin{figure}
	\bc
	\resizebox{.7\columnwidth}{!}{%
		\includegraphics{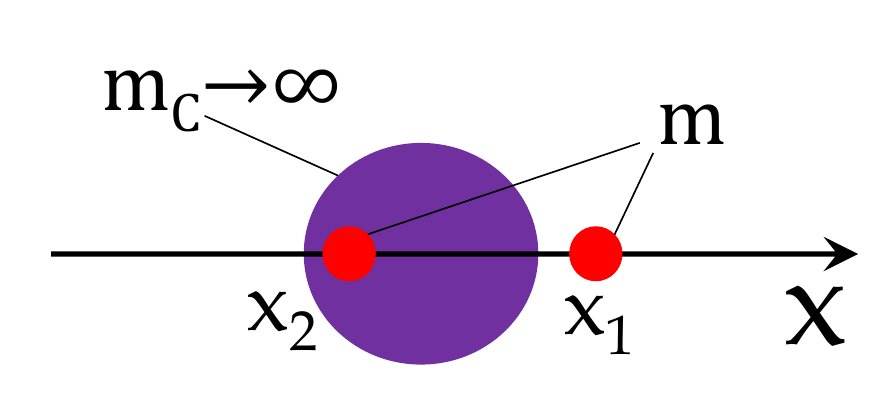}}\ec
	\caption{One-dimensional three-body model utilized in Ref. \cite{2018OI03}. 
		The core nucleus is assumed to be infinitely heavy.}
	\label{fig:Oishi_01}       
\end{figure}

In Ref. \cite{2018OI03}, the TD calculation has been applied to the 
one-dimensional system, which consists of two fermions and one 
core (daughter) nucleus. 
Fig. \ref{fig:Oishi_01} visually displays this system. 
This three-body system can be a good testing field to discuss 
the 2F emission, where the core nucleus works as the source 
of the potential barrier $V_C(x)$. 
Thus, the corresponding Hamiltonian reads 
\beqa
\oprt{H} &=& \hat{h}(1) + \hat{h}(2) + v_{12} (x_1 -x_2), \nonumber \\
h(i) &=& -\frac{\hbar^2}{2m} \frac{d^2}{dx^2_i} + V_C(x_i), 
\eeqa
where $m$ is the single-fermion mass. 
For the core nucleus, an infinitely-heavy mass is assumed.

\begin{figure}
		\bc
	\resizebox{.7\columnwidth}{!}{%
		\includegraphics{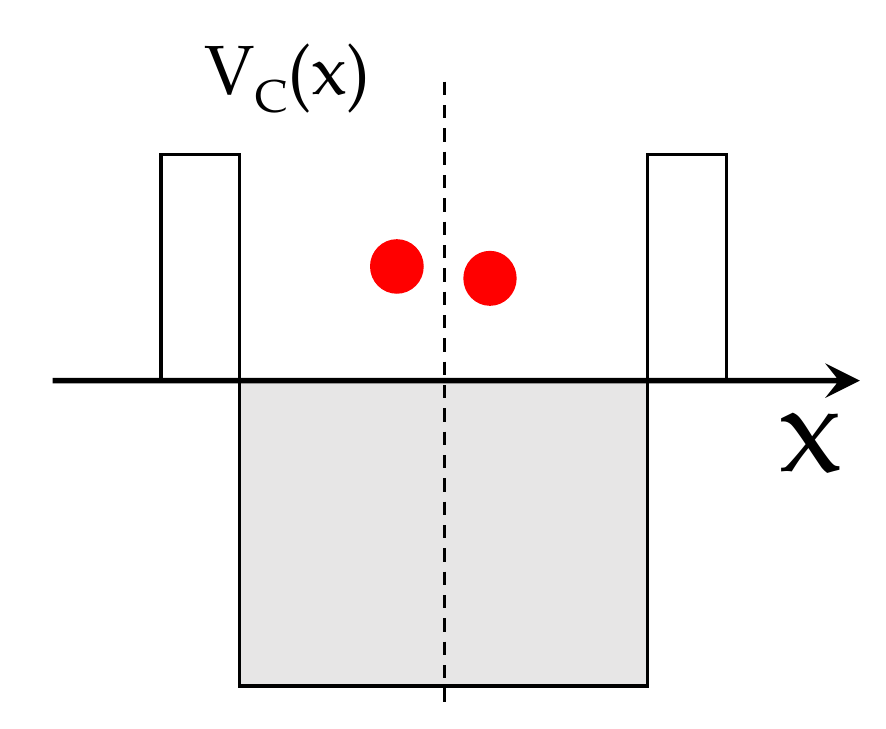}}\ec
	\caption{Two fermions confined inside the potential $V_C(x)$ at $t=0$. }
	\label{fig:Oishi_02}
\end{figure}

For the pairing correlation of the emitted fermions, 
in Ref. \cite{2018OI03}, the short-range attraction 
$v_{12} (x_1 -x_2)$ is employed. 
The core-fermion potential $V_C(x_i)$, on the other side, 
includes the repulsive-barrier and the attractive-well 
terms, consistently to the realistic radioactive systems. 
See Fig. \ref{fig:Oishi_02} for a schematic picture of $V_C(x)$. 
For numerical calculations, continuum states are discretized. 
Thus, Eq. (\ref{eq:init_c}) is modified 
as $\ket{\psi_0} = \sum_{i} \mu_i \ket{E_i}$, 
where the eigen-energy $E_i$ is obtained by diagonalizing 
the Hamiltonian within the box-discretized model space.

In order to fix the initial state $\ket{\psi_0}$, 
the confining-potential procedure has 
been utilized in Ref. \cite{2018OI03}. 
Namely, at $t=0$, the wave function of two fermions is 
solved so as to be localized inside the potential barrier. 
Figure \ref{fig:Oishi_02} schematically shows this situation.

By monitoring the time evolution of the 2F state, 
remarkable features of the 2F-tunneling process 
have been found \cite{2018OI03}, that is here summarized: 
\begin{itemize}
	\item Pairing-dependence of the tunneling probability or decaying lifetime. 
	Starting from the confined 2F state, the survival probability 
	$P_{\rm surv}(t)$ is indeed well approximated by the exponential 
	function as in Eq. (\ref{eq:EXP_PS}). 
	Furthermore, as long as in the weak-pairing region, the mean lifetime $\tau$ 
	is extended when the attractive pairing is switched on. 
	This means that the tunneling probability via the potential 
	barrier can be reduced by the pairing attraction. 
	This result can be explained from the kinematics. 
	Because of the attractive pairing, the 2F-energy level is decreased. 
	This lower 2F energy with respect to the fixed barrier height 
	then leads to a smaller tunneling probability. 
	\item Correlated emission. 
	When the total spin of 2F is coupled to zero ($S_{12}=0$), 
	with the attractive-pairing force, 
	there can be a spatially-localized component of the 2F-wave function 
	during the tunneling process. 
	Or intuitively, two fermions are promoted to 
	(i) move to the same direction, and (ii) be emitted at the same time. 
	This is a typical product of the pairing correlation in the continuum. 
	Note also that the similar localization has been predicted for the 
	bound systems, as the dinucleon correlation \cite{2011Ha01}. 
	On the other side, when the pairing correlation is switched off, the
	so-called sequential emission becomes the dominant process: 
	only one fermion is emitted whereas the other one is left 
	inside the potential. 
	\item Total-spin dependence. 
	In the $S_{12}=0$ case, the 2F emission can be affected by the 
	pairing correlation as mentioned above. 
	On the other side, when two fermions are coupled to $S_{12}=1$, 
	the pairing effect vanishes: the results with and without 
	the pairing force are not changed when $S_{12}=1$. 
	Furthermore, the $S_{12}=1$ emission can be faster than the 
	$S_{12}=0$ emission, even if the effective barrier height is the same. 
	This total-spin dependence is indeed a product of the Pauli principle, 
	which gives the additional constraint on the spatial part of the 
	wave function. 
\end{itemize}
Note that there have been other TD calculations performed for 
the three-dimensional 2F tunneling \cite{2014OI02}, 
where similar conclusions are obtained. 
Further developments of TD calculation are expected to lead to a deeper 
knowledge of the multi-fermion dynamics and of
pairing correlation in the continuum.

\section{Two-neutron pairing correlations} \label{TNPC}
Closely tied with the previous topic is the role the role 
of the pairing correlations in two-neutron transfer reactions \cite{1989LO01,2001VO20}. Of course one-dimensional models are nice tools, but ultimately one wants a full three-dimensional picture of the pairing correlations and of the transfer mechanims. Vitturi has investigated several aspects of the delicate interplay between correlations and dynamics, in collaboration with many, some of which are collected in the following sections.

\begin{figure*}[tb]
	\begin{center}
		\resizebox{.95\textwidth}{!}{
			\includegraphics{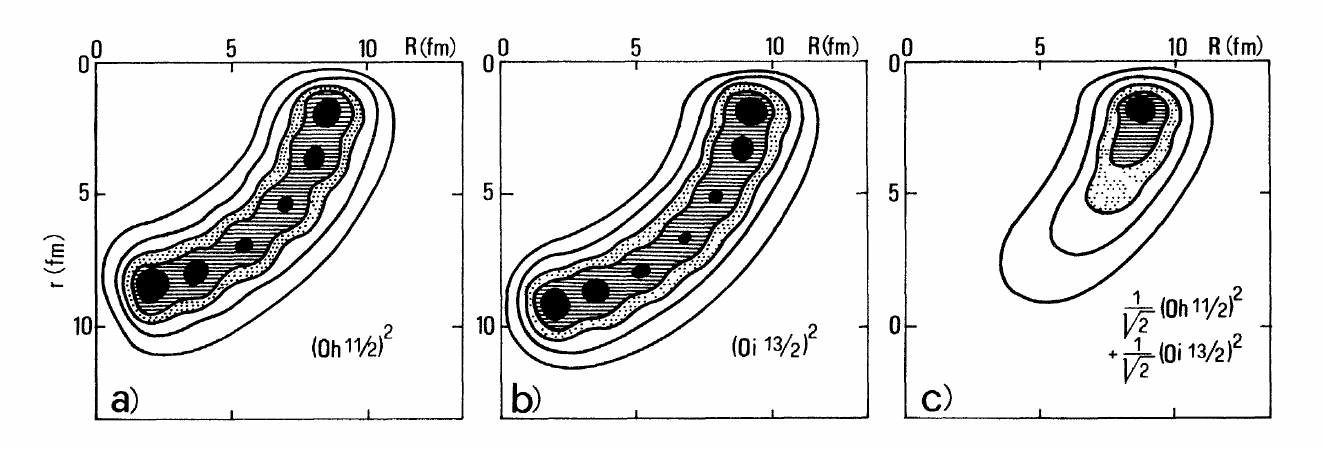}		}
	\end{center}
	\caption{ Two-particle densities as a function of the relative distance, $\vec{r}$, 
		and the center of mass coordinate, $\vec{R}$, 
		of two particles inside a nucleus. 
		These are obtained with simple harmonic oscillator wave functions, 
		assuming particular configurations indicated in the figure. 
		Taken from Ref. \protect\cite{1984CA04}. }
	\label{fig:dineutron}      
\end{figure*}

In this connection, Vitturi's contribution to the problem of 
dineutron correlation should be emphasized. The dineutron correlation 
is a spatial correlation due to the pairing, with which two neutrons inside a 
nucleus are strongly localized in space. It has attracted lots 
of attention in recent years in connection to exotic nuclei \cite{2005HA59}. 
It should be emphasized that the paper by Vitturi, together with Catara, Insolia, and 
Maglione \cite{1984CA04}, was the first one which pointed out that the dineutron correlation is caused by admixtures of a few single-particle orbits with opposite parity. 
Fig. \ref{fig:dineutron} shows the two-particle density, 
obtained with simple harmonic oscillator wave functions,  
as a function of the relative distance between two neutrons, 
$\vec{r}=\vec{r}_1-\vec{r}_2$, where $\vec{r}_1$ 
and $\vec{r}_2$ are the coordinates of each neutrons, 
and the center of mass of the two neutrons, $\vec{R}=(\vec{r}_1+\vec{r}_2)/2$.  
This figure clearly shows that the two-particle density is well localized 
in the small $r$ region, if two configurations with single-particles orbits 
with opposite parities, namely $(0h_{11/2})^2$ and $(0i_{13/2})^2$, are mixed in the wave function, whereas the pure configurations lead to an extended density distribution. 
Such surface clustering of two-neutrons due to the dineutron 
correlation plays an important role in two-neutron transfer 
reactions \cite{1986BE08,1988LO05,1989IN01}, 
which are discussed in the next section. 

\section{Two-neutron transfer mechanism}\label{2N}

Two-neutron transfer cross sections are the probe {\it par excellence} for testing pairing correlations in nuclei. Ground state to ground state $^{A}$X($p$,$t$)$^{A+2}$X or $^{A}$X($t$,$p$)$^{A+2}$X cross sections are enhanced by the presence of pairing with respect to the case of pure configurations of two nucleons in single-particle orbitals. From the structure point of view, pairing produces a coherent admixture of different configurations in the ground state, while higher lying $0^+$ states are non-collective with respect to the pairing operator. During the transfer reaction to the ground state, all these configurations contribute constructively to the total transfer cross section, thus producing the expected enhancement.

On the other hand, one would like this probe to be sensitive to each contribution, i.e. to the spectroscopic factor for each configuration. One has one datum versus multiple different possibilities. In a large open shell, one can easily have five or six components. The same enhancement could be obtained in principle with different mixing of the  different contributions.  It is known that certain orbitals contribute less than others to the transfer cross section since the two neutrons  barely couple to a $0s$ state in the relative motion coordinate. However, this will only help in very few cases. On top of that, one has to face other uncertainties like the optical potentials needed to calculate total cross sections.

In order to overcome this issue, the idea was to complement ($p$,$t$) and ($t$,$p$) cross sections with two-neutron transfer with heavier probes. Ratio between the cross section to the ground state and to the first excited 0$^{+}$ states of $^{112}$Sn, $^{32}$Mg, and $^{68}$Ni were studied in~\cite{2014La08} using ($t$,$p$), ($^{18}$O,$^{16}$O), and ($^{14}$C,$^{12}$C) reactions. The original calculations in~\cite{2014La08} were done including only the simultaneous transfer of the neutrons in Zero Range Distorted Wave Born Approximation (DWBA) for ($t$,$p$), performed with DWUCK~\cite{DWUCK}, and in second-order perturbation theory along a semiclassical trajectory for the heavier probes, done with the code TFF~\cite{2017Fo01}.

\begin{figure*}\bc 
	\resizebox{1.4\columnwidth}{!}{%
		\includegraphics{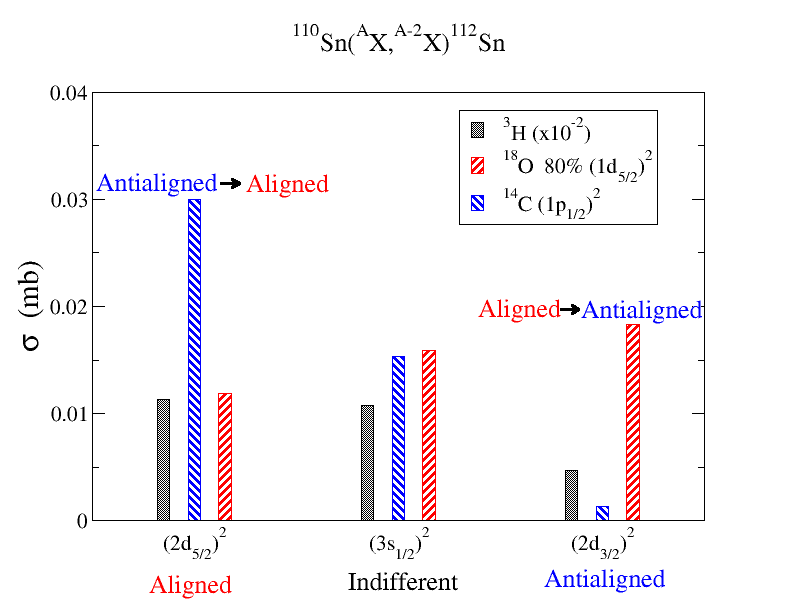}
	}\ec
	\caption{Comparison of ($t$,$p$), ($^{18}$O, $^{16}$O), and ($^{14}$C,$^{12}$C) total two-neutron transfer cross sections for different independent particle configurations of the two neutrons in $^{112}$Sn. }
	\label{fig:histog}       
\end{figure*}

By comparing the results with the different probes applied to the first nucleus of interest, $^{112}$Sn, one finds an interesting selectivity with respect to the alignment of the spin and the angular momentum of each neutron. In order to illustrate this effect, Fig.~\ref{fig:histog} shows the cross sections for ($t$,$p$), ($^{18}$O,$^{16}$O), and ($^{14}$C,$^{12}$C) reactions assuming different pure configurations for the structure of $^{112}$Sn. Three different pure configurations were chosen: one with zero orbital angular momentum, ($s_{1/2}$)$^2$, one with the spin and angular momentum aligned ($d_{5/2}$)$^2$, and the final one with spin and angular momentum antialigned ($d_{3/2}$)$^2$. The first important thing to notice is that total cross sections obtained for ($t$,$p$) barely change from one configuration to the other. Neutrons in the triton are initially in a $s-$wave and thus they are not sensitive to the alignment of the spin in their final configuration after the reaction. However, those neutrons with an aligned initial configuration give a larger cross section when the final configuration is antialigned and vice-versa. One can see in Fig.~\ref{fig:histog} that $^{14}$C gives a higher cross section when the configuration in $^{112}$Sn is aligned, ($d_{5/2}$)$^2$, whereas $^{18}$O gives a higher cross section for the antialigned one, ($d_{3/2}$)$^2$.

These configurations will be mixed in the actual total experimental cross section, but the enhancement produced for the three probes will be completely different. Such difference will give information about the relative weight of aligned and antialigned configurations, i. e. the spectroscopic factors for the nucleus of interest. Moreover, within the same reaction one can often measure also the total cross section to the first excited state of the target. The ratio between the cross section to the first 0$^+$ excited state and the ground state will be sensitive to this microscopic information, thus being qualitatively different for each probe. The added value of looking at the ratios is that the dependence on the optical potentials will be reduced.

\begin{figure}
	\resizebox{1.\columnwidth}{!}{\includegraphics{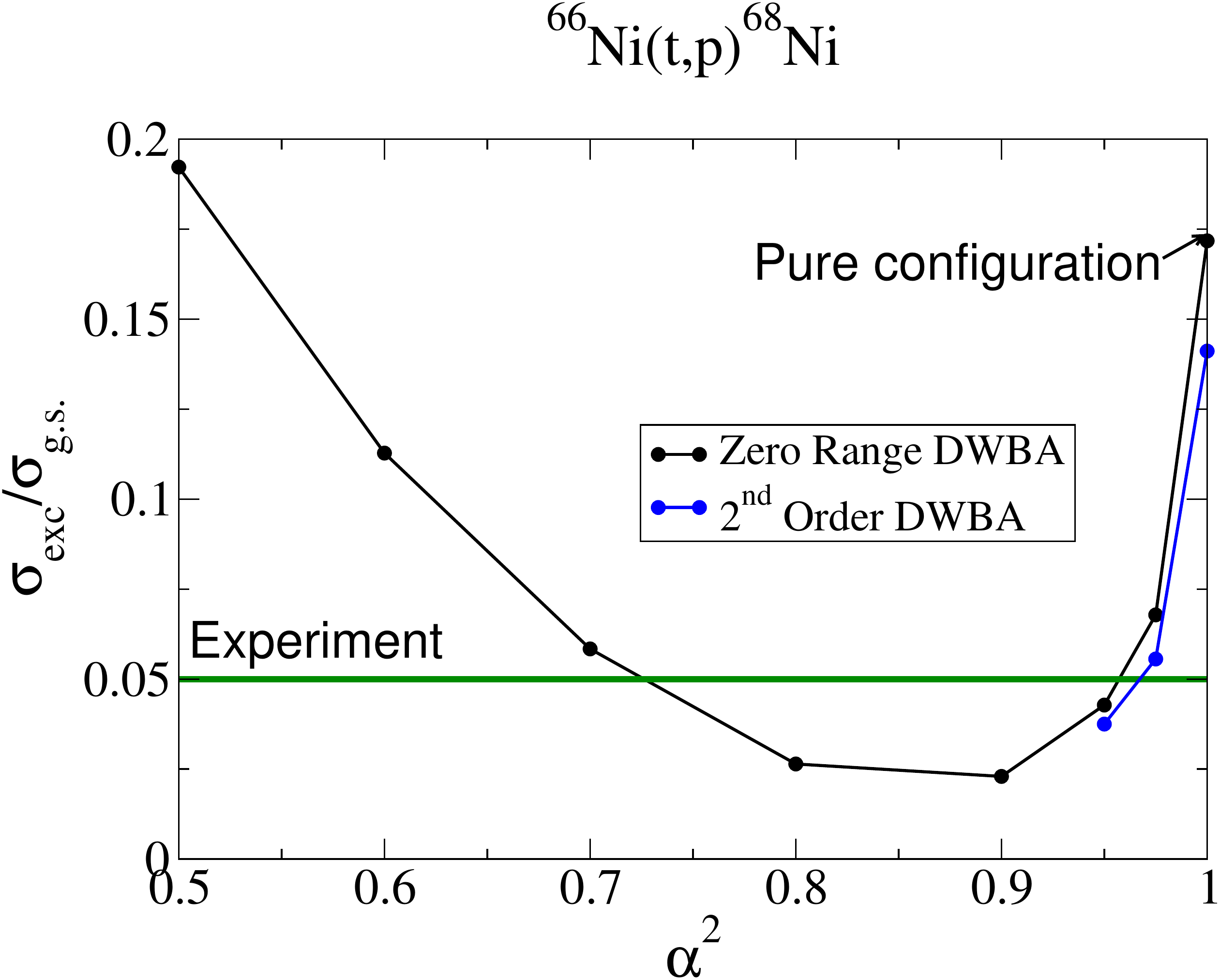} }
	\caption{Ratio between the total cross section to the first 0$^+$ excited state and to the ground state for the $^{66}$Ni($t$, $p$)$^{68}$Ni two-neutron transfer as a function of $\alpha^2$.  The results around 1 are quite independent on the particular scheme used to calculate cross-sections.}
	\label{fig:ratio}       
\end{figure}

In Fig.~\ref{fig:ratio}, we show the ratio between the total cross section to the first 0$^+$ excited state and the ground state for the $^{66}$Ni($t$,$p$)$^{68}$Ni case. This ratio varies with the parameter $\alpha$ which is proportional to the mixing of 0p-0h and 2p-2h configurations. More precisely the model wavefunctions of the ground state and first excited state used here are of the form

\begin{eqnarray}
\mid 0^+_{gs} \rangle &~=~&\alpha \mid 0 \rangle+\beta~~\mid (g_{9/2})^{2}(p_{1/2})^{-2} \rangle \\
\mid 0^+_{exc} \rangle &~=~&-\beta\mid 0 \rangle+\alpha~~\mid (g_{9/2})^{2}(p_{1/2})^{-2} \rangle
\end{eqnarray}
where $\mid 0 \rangle$ corresponds to the filling of the four lowest major neutron shells. So, $\alpha$ will be one if $N=40$ at $^{68}$Ni is a magic number. To test how robust this observable is,  the ratio in Zero Range \cite{DWUCK} and in full second order DWBA \cite{Thomp} has been calculated, which includes finite range successive, simultaneous and non-orthogonality terms. This latter calculation was performed with the code FRESCO~\cite{1988Th01}. We see in Fig.~\ref{fig:ratio} that both calculations will give raise to the same conclusion even though they can give very different values for each cross section. In this case, this ratio is only compatible with the experimental one \cite{Rab-Els,2019FL06} \footnote{In the recent Ref. \cite{2019FL06}, a more elaborate mixing of different configurations is proposed to explain the angular distribution.}, for a large value of $\alpha^2\approx 0.95$ showing that it is very close to a shell closure situation. An $\alpha^2\approx 0.7$ is also consistent with the experimental data. This alternative could be ruled out if one could compare with the experimental ratio for the ($^{14}$C,$^{12}$C) case. The antialigned neutrons in the Carbon probe will give a very  small cross section going to the  $p_{1/2}$ orbit compared to the $g_{9/2}$ one, giving rise to a very different dependence of the ratio on the mixing coefficient $\alpha^2$.

In conclusion, these ratios are very robust observables which help to constrain the microscopic information of the nuclei of interest. In addition, the combination of  different probes, light versus heavy projectiles, allows to disentangle different possibilities.

Another interesting case where two-neutron transfer process could be of the utmost importance is that of shape-phase transitions in nuclei.  As shown in~\cite{2007Fo08,2017ZH13}, the total two-neutron transfer cross sections along an isotopic chain varies strongly when a shape-phase transition is passed. At present, some of us, in collaboration with A. Vitturi, are currently investigating two-neutron transfer processes as a tool to distinguish between standard shape-phase transitions (i.e. first or second order phase transitions occuring in systems as a function of a certain control parameter) and situations where the phase transition is driven by shape coexistence ~\cite{2018Vi01,2019La01}.

\section{Pairing in the continuum} \label{PC}

Regarding pairing in Nuclear Physics, an interesting open problem deals with the complications arising in its treatment and behaviour in the continuum. Traditionally, studies of pairing have been done having in mind heavy stable nuclei. In this region of the Segrè Chart, the neutron separation threshold is large so that the main single particle states contributing to the pairing are strongly bound.

The development of radioactive ion beam facilities all around the world are opening access to a great deal of information on regions of isotopes closer to the neutron drip line at larger masses than ever before. These new isotopes present the same fingerprints of pairing as in stable nuclei, but now the single particle states involved are weakly bound or directly unbound, i.e. in the continuum.

Therefore, a proper treatment of the continuum is compulsory for future studies of pairing close to the drip line. To this end, the standard BCS approach fails to reproduce the spatial asymptotic properties of the ground state of the nuclei studied. It produces an unphysical gas of neutrons surrounding the nucleus~\cite{1996Do06}. This problem has pushed the community toward more complex models like Hartree-Fock-Bogoliubov (HFB) \cite{RS book}, Gamow Shell Model (GSM) \cite{2002Mi00} or the Configuration-Space Monte Carlo (CSMC) method \cite{1993Ce}. However, the hardness and complexity of such calculations prevents from systematic calculations to study the sensitivity of pairing to the particular structure of the continuum or to the proximity of the Fermi level to the neutron separation threshold.

One of the few existing approaches to solve the problem of BCS in the continuum is the use of the continuum Single Particle Level Densities (CSPLD) by R. Id Betan and collaborators~\cite{2012Be01}. A different approach has been proposed that relies on discretizing the continuum with the use of a Transformed Harmonic Oscillator (THO) basis~\cite{2016La17}. The combined versatility of the THO basis and the BCS approach allowed to study the behaviour of the pairing in the presence of resonances or weakly bound states and the evolution of different  properties of the nuclei when making the system less and less bound. At the same time the results from Ref. \cite{2012Be01} were also reproduced.

The main result found in~\cite{2016La17} was an enhancement of the occupation of the low-lying continuum when a weakly bound single particle state is present. This occupation is not exclusive of $s-$waves like in halo nuclei but it occurs in the same total and orbital angular momenta as the weakly bound state. In Ref.~\cite{2016La17} it was shown that this occupation increases as long as the bound state becomes less and less bound. This effect was already seen in a more complex HFB calculation~\cite{2001Gr32}, but the lack of flexibility made it difficult to study the nature of this occupation more in depth.

\section{Dynamics of two-neutron transfer reactions} \label{D2nTR}
The transfer reactions are known to have a high selectivity. For this reason, they can be used to obtain spectroscopic information of the interacting nuclei. The one-particle transfer reaction put in evidence the single-particle component of the nuclear wave functions. The two-particle transfers, on the other hand, are usually used to study the particle correlation between the two transferred particles. One crucial point, in this case, is to answer the question if these two particles are transferred sequentially (one after the other), or at the same time (in a one-step process). The transfer of heavier particles (like tritons, $\alpha$ particles, etc) is connected with the cluster components of the wave functions. These components are usually relevant in nuclei that are multiple of the $\alpha$ particle, like, $^8$Be, $^{12}$C, $^{16}$O, etc. or close to that. Reviews and textbooks on this line have been published \cite{2001VO20,PIB13,KBa77,Sat83}. 

The transfer reactions are also an essential method of production of new nuclear species. A good example of this is the production of some radioactive beams as a secondary beam in the reactions $^9$Be($^7$Li,$^6$He), $^9$Be($^7$Li,$^8$Li), $^3$He($^6$Li,$^8$B), and many others. 

\begin{figure}
	\resizebox{1.\columnwidth}{!}{\includegraphics{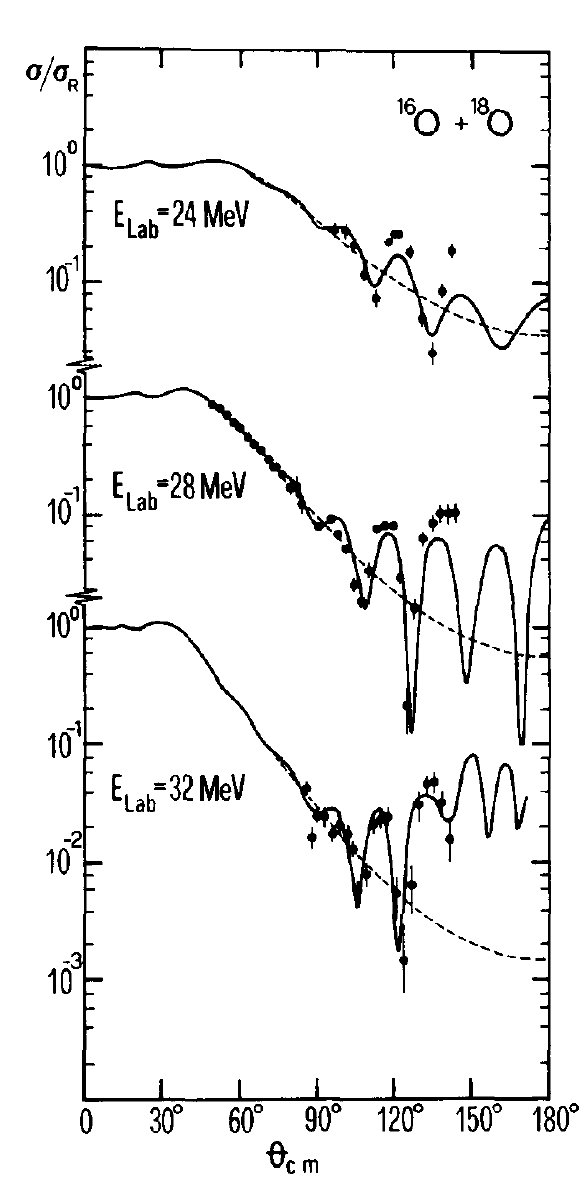}}
	\caption{Comparison of the elastic scattering angular distributions considering pure elastic reaction (dashed line) and the interference between elastic and elastic transfer processes (full line). Figure was taken from Ref.~\protect\cite{VDa86}.}
	\label{fig1}       
\end{figure}

One important class of transfer reactions that has obtained special attention is the elastic transfer. In this reaction, the initial and final partitions are identical, and this might come as a result of elastic scattering or multiple (at least two) back-and-forth transfer processes. The final nuclei are indistinguishable. This makes a necessity to sum the amplitudes of both processes coherently, i.e. the theoretical cross sections are calculated by considering 
\be
\frac{d \sigma}{d \Omega} \propto\vert T_{scatt}(\theta) + T_{tr}(\pi - \theta)\vert ^{2} \;,
\ee
where  $T_{scatt}$ and $T_{tr}$ correspond to the amplitudes of the elastic scattering and elastic transfer channels, respectively. This is a very convenient reaction for studying the structural characteristics of some nuclei. The reason is that one usually needs the spectroscopic amplitudes for two (projectile and target) overlaps in order to derive the transfer reaction cross section. In the elastic transfer, these two overlaps coincide, and this simplifies the structure calculations, once the
interference of the elastic and transfer amplitudes is considered. The interference of the two contributions can give rise to oscillatory patterns, that might be significant depending on the considered bombarding energy and angle. This behavior has been confirmed by experiments, as for example in the $^{12}$C + $^{13}$C and $^{18}$O + $^{16}$O reactions at low energy \cite{VDa86}. In Fig.~\ref{fig1} the effect of the interference of the elastic scattering and elastic transfer amplitudes is shown for the $^{18}$O + $^{16}$O reaction at E$_\mathrm{Lab}$ = 24, 28, and 32 MeV. The dashed line corresponds to the results when only the elastic scattering amplitude is used to determine d$\sigma / \mathrm{d}\sigma_\mathrm{R}$. When the elastic transfer amplitude  is included (full line) a strong oscillatory pattern is observed at backward angles. The same result was also shown for the $^{12}$C + $^{13}$C reaction.
Recently studied examples of the elastic and elastic transfer interference are the two-neutron transfer reactions of $^{16}$O($^{18}$O, $^{16}$O)$^{18}$O \cite{ECL16} and $^9$Be($^7$Be, $^9$Be)$^7$Be \cite{UPL19}, and the alpha transfer reaction $^{16}$O($^{12}$C, $^{16}$O)$^{12}$C \cite{FLL19}. 

In the case of the two-neutron transfer mentioned above, one important question to understand the details of the dynamic of the nuclear reaction is whether the two neutrons are transferred in one step or two steps. We emphasize that we are concerned with the transfer process itself. So, the projectile or the target might be excited before the transfer process occurs, and we are not considering this process as one of the steps. The two-neutron transfer were extensively used in the past using the ($t,p$) reactions. The reaction was quite simple in this case due to the simplicity of the projectile overlaps that involves a small model space to describe the wave functions of the projectile and the ejectile. At present, this kind of reactions is prohibited in several labs due to safety reasons related to tritium. A natural candidate to study the two-neutron rearrangement is the ($^{18}$O,$^{16}$O) reaction because the $^{18}$O can, to a good approximation, be considered as an inert $^{16}$O core plus two neutrons, mainly in the $spd$ shell. This kind of reactions is a powerful tool to study the effect of pairing correlation in the two-neutron transfers reactions.

Various recent studies on the effect of pairing correlations in two-neutron transfer reactions to the ground, as well as, to some excited states of the residual nuclei have been performed: $^{12,13}$C($^{18}$O, $^{16}$O)$^{14,15}$C \cite{CCB13,2015CA01,CJC17}, $^{16}$O($^{18}$O, $^{16}$O)$^{18}$O \cite{ECL16}, $^{28}$Si($^{18}$O, $^{16}$O)$^{30}$Si \cite{CLL18},   $^{64}$Ni($^{18}$O, $^{16}$O)$^{66}$Ni \cite{PSV17},  and $^{206}$Pb($^{18}$O, $^{16}$O)$^{208}$Pb \cite{PSR15}. In all of these studies a finite-range coupled reaction channel method was used to calculate the direct two-neutron reactions, while the two-step coupled channel Born approximation was used for the two-step processes. To determine the spectroscopic amplitudes extensive shell model calculations were performed in most of the cases. For Nickel isotopes Interacting Boson Model-2 (for even isotopes) and Interacting Boson Fermion Model (for odd isotopes) structure calculations were also performed. These works concluded that the one-step (direct) two-neutron transfer prevails over the two-step (or sequential) processes for the light systems \cite{CCB13,CJC17,ECL16}, showing the relevance of the pairing correlations in the two-neutron transfer in this case. For the reaction of medium-light $^{28}$Si and medium mass target $^{64}$Ni, a strong competition between the pairing and collective residual interaction was observed. For the ground state of $^{66}$Ni (that is almost spherical, and consequently with low collectivity)  a predominance of the direct two-neutron transfer was observed, while for the first excited states of $^{66}$Ni, as well as for the ground states and several excited states of $^{30}$Si, that has stronger degree of collectivity, the two-step process dominates the reaction mechanism.

\begin{figure}
	\resizebox{1.\columnwidth}{!}{\includegraphics{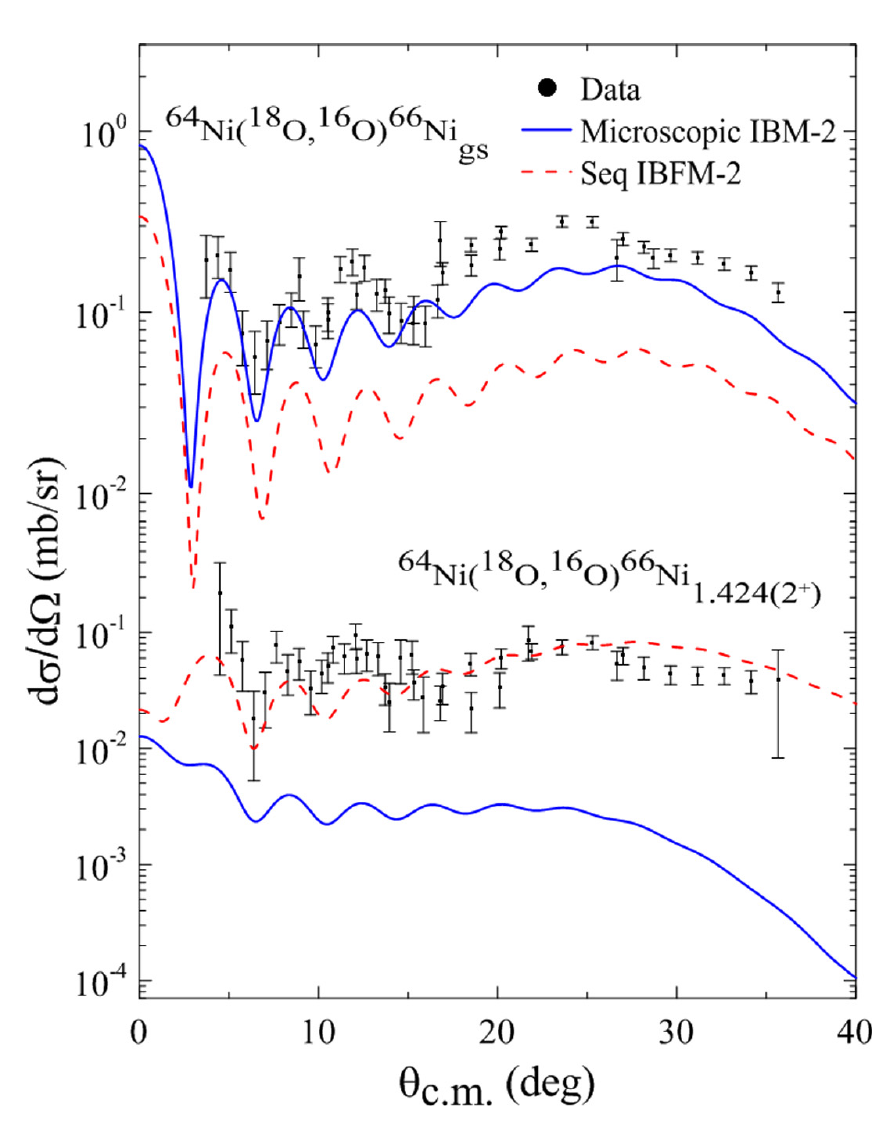}}
	\caption{Comparison of the experimental angular distributions with one- (full line)  and two-step (dashed line)  transfer theoretical cross section using spectroscopic amplitudes derived from Interacting Boson Model calculations. Figure taken from Ref.~\protect\cite{PSV17}.}
	\label{fig2}       
\end{figure}

The results for the $^{64}$Ni($^{18}$O,$^{16}$O)$^{66}$Ni \cite{PSV17} are shown in Fig.~\ref{fig2} where various versions of the Interacting Boson Model were used to derive the spectroscopic amplitudes needed in coupled channel calculations. It is important to notice that in all these calculations it was not necessary to use any unhappiness factor \cite{Thomp,1979VAA} to describe the correct order of magnitude as it was often the case in the past for certain ($t,p$) transfer reactions. 

In the case of the heavier system, the two-neutron transfer reaction $^{206}$Pb($^{18}$O,$^{16}$O)$^{208}$Pb is dominated by the two-step process, showing that the effect  of pairing correlations in two-neutron transfer reactions depends not only on the structure of the reacting nuclei, but also on their masses.

\section{Direct reactions at intermediate energies in the eikonal approximation}\label{eikonal}
\subsection{Applications based on the Glauber model}
Nuclear reactions induced by heavy ions at intermediate energies give rise to a variety of phenomena, such as elastic and inelastic scattering, transfer and charge-exchange reactions.  While at low bombarding energies, the scattering process and the internal degrees of freedom of the colliding nuclei are strongly mingled, they can be decoupled in reactions induced by light ions at high energy. 
One of the research lines conducted by Andrea Vitturi and collaborators was based on the description of these reactions in a microscopic approach. The theoretical model is a simplified version of the multiple scattering theory of Goldberger-Watson, originally deduced by Glauber~\cite{Glauber}. In the optical limit of this model, nucleus-nucleus processes are microscopically described in terms of nucleon-nucleon collisions and each partial wave phase shift is obtained considering straight-line trajectories~\cite{GM2}.  Andrea Vitturi and Francesco Zardi successfully modified this formalism to the description of the elastic scattering at intermediate energies between heavy nuclei~\cite{VZ86} by replacing the straight line by the Rutherford trajectory that takes into account the strong Coulomb repulsion when heavy ions are involved. The relative motion is reliably described in the eikonal approximation in terms of a phase shift derived from elementary interactions and nuclear densities, taken from the current phenomenology with no fitting procedure. In particular, there is no ambiguity associated with the choice of an optical potential.  The agreement of these calculations with the experimental data was very successful and encouraged them to extend the model to the description of other nuclear reactions.

In Ref.~\cite{LVZ1} the formalism was derived for inelastic scattering processes. Two different approaches were explored. The ``standard'' one considered that the target is excited in projectile-nucleon collisions where contributions from multiple scattering processes are preliminary summed to produce the projectile-nucleon elastic scattering amplitudes. Alternatively, the authors introduced a fully microscopical approach where the inelastic process is produced by single nucleon-nucleon collisions. In this case, the inelastic transition amplitudes were obtained by integrating the nuclear densities and transition densities together with the experimental nucleon-nucleon scattering amplitudes. With no free parameters at all, they succeeded to describe a large variety of elastic and inelastic reactions at intermediate energies (30-350 MeV/A)~\cite{LVZ2}. 

The ensuing extension of the formalism to processes that involve several coupled channels is very interesting. The method, introduced in Ref.~\cite{LZV1}, avoids several difficulties of multi-scattering theories, in particular, the low convergence of the series.  Indeed, the set of coupled equations for the S-matrix elements are replaced by simple matrix relations that involve the basic ingredients of the Glauber model, i.e. the densities and transition densities and the nucleon-nucleon scattering amplitude. In some particular cases the summed series led to analytic expressions.
Mutual excitation processes, such as charge-exchange reactions can be studied as an extension of this method. In ref.~\cite{MLVZ} the formalism has been developed and applied to charge-exchange reactions between 50 and 140 MeV/A. The overall agreement of the quantities calculated within this approximation with the available data was very good, in spite of the many complications that might be hidden in the fact that the cross sections arise from the coherent contribution of many different components both in the nuclear transition densities and in the nucleon-nucleon interaction.

\begin{figure*}[t!]\bc
	\resizebox{1.5\columnwidth}{!}{\includegraphics{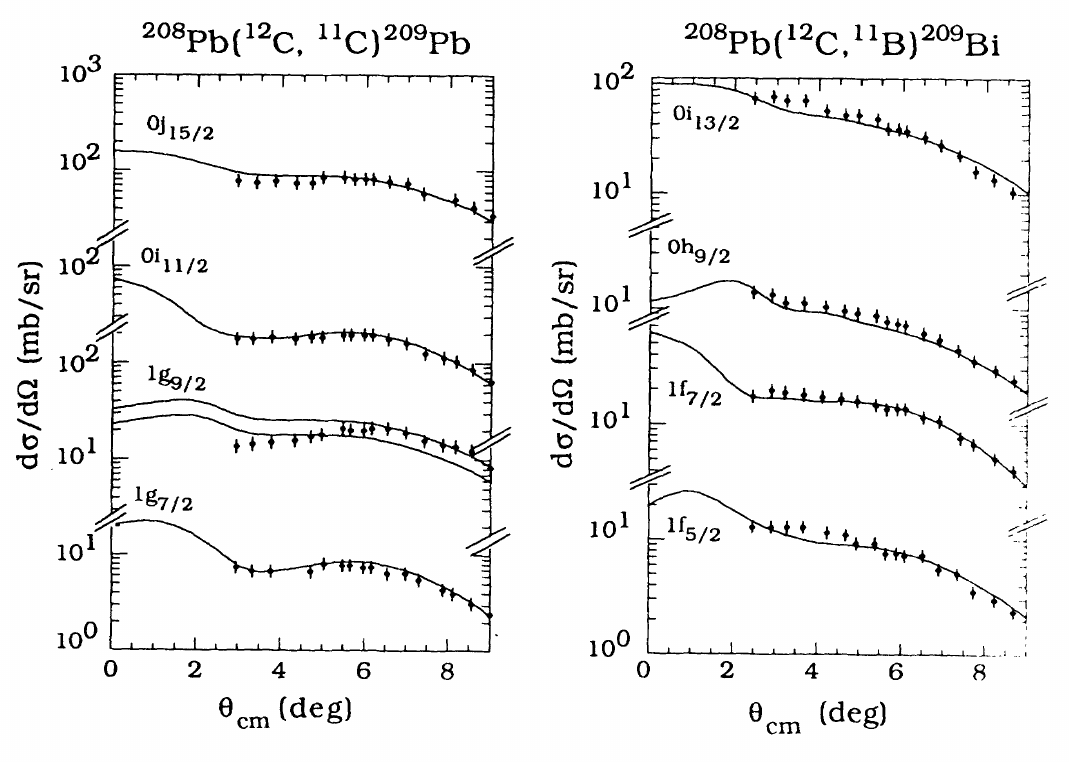}}\ec
	\caption{Angular distributions for the single-neutron and single-proton transfer reactions induced by $^{12}$C on $^{208}$Pb at 480 MeV. Taken from ref.~\cite{BLVZ}.}
	\label{fig:eiko}     
\end{figure*}

Reactions that involve the transfer of nucleons at intermediate energy between heavy ions have also been studied with the Glauber model. In a first work, multi-pair transfer processes have been addressed.  Combining a macroscopic approach for the pairing structure part with the sudden limit of the eikonal approximation for the description of the scattering process, quantitative predictions for  
multi-pair transfer cross sections at different bombarding energies were obtained~\cite{LZV2}. In a second work, one-nucleon transfer reactions were addressed in a Glauber-like approach and compared with exact finite-range distorted wave approximation calculations. A better agreement with data was observed for the former~\cite{BLVZ}. These results are reported in Fig. \ref{fig:eiko} for reactions induced by $^{12}$C on $^{208}$Pb (a neutron and a proton transfer respectively).

While in  heavy-ion reactions at energies around the Coulomb barrier both nuclear and Coulomb excitations are treated on the same footing, this was not the case at high energies, where the studies concentrated on one or the other process. To put in evidence the interference between the two interactions, a new formalism has been developed, based on the eikonal approximation, that allows to treat both processes from a unified perspective~\cite{ZLV}.

In conclusion, in these pioneering works A. Vitturi and collaborators have shown the reliability of formalisms {\em\`a la Glauber} for the description of a large variety of reactions between heavy ions at intermediate energies in a parameter-free approach.

\section{Low-lying strength and clustering in weakly bound nuclei} \label{Cont-Clus}
The advent of nuclear reactions with radioactive beams has triggered a significant focus on the exotic properties of nuclei close to (or even beyond) the drip-lines~\cite{2003HA50}. The small binding energy of light neutron-rich systems, such as $^{11}$Li or $^{12}$Be, was soon associated to the abnormal density distribution of the valence neutrons extending way outside the mean field generated by the remaining core and gave rise to the very concept of nuclear halo~\cite{1985TA18,2013TA17}. Studies along these lines evidenced unusual concentrations of strength appearing at low excitation energies in photo-excitation or break-up processes (e.g.,~\cite{1993IE01}).

\begin{figure}
	\resizebox{1.02\columnwidth}{!}{%
		\includegraphics{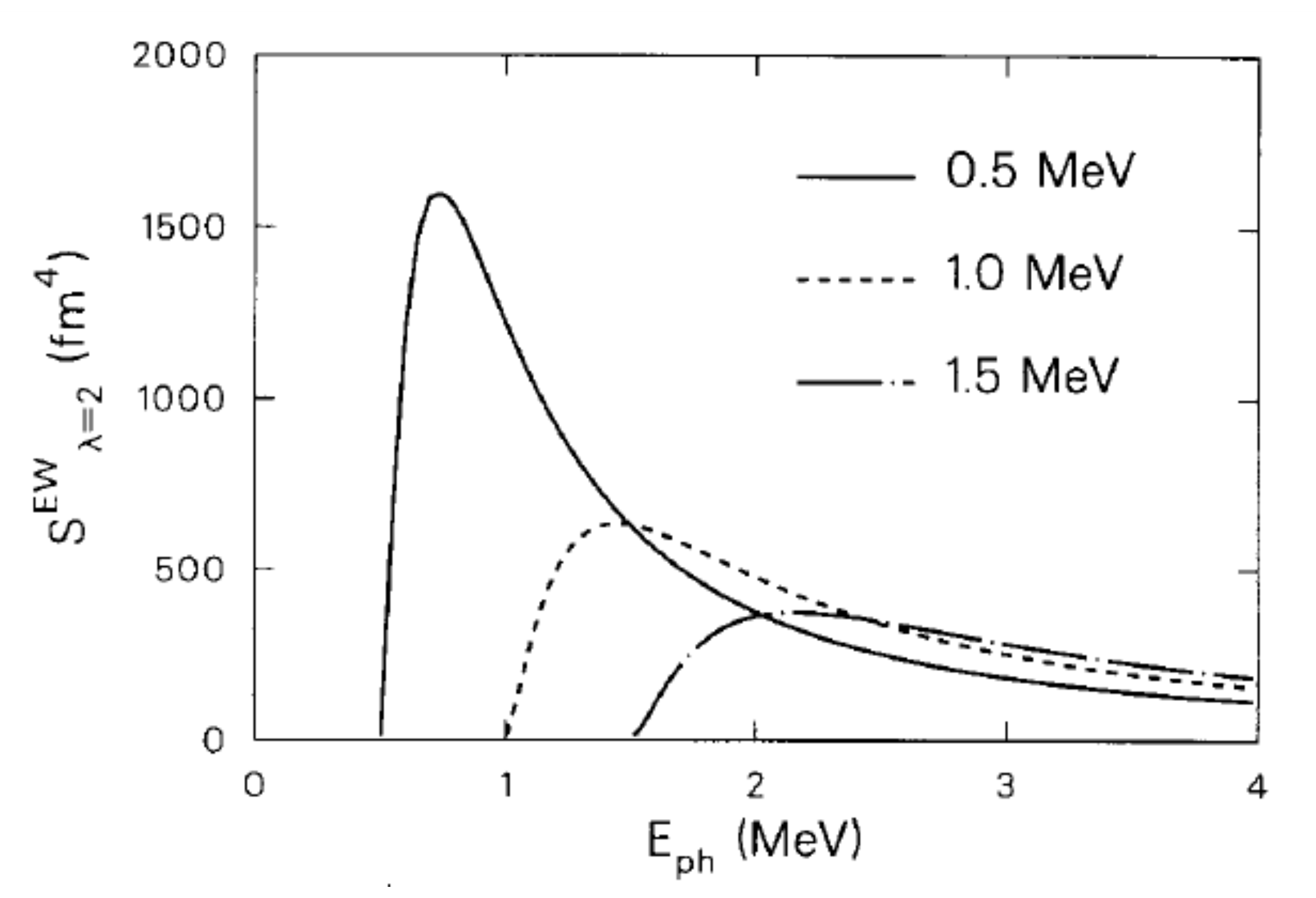}
	}
	\caption{Energy-weighted quadrupole strength for transitions from a $2p$ weakly bound orbital into the $p$-wave continuum for three different binding energies 0.5, 1.0 and 1.5 MeV. The peaks shift towards higher energies, while the total strength decreases, as the binding energy increases. Figure adapted from Ref.~\cite{1996Ca21} with the authors' permission.}
	\label{fig:binding}       
\end{figure}

The pioneering work of Catara, Dasso and Vitturi~\cite{1996Ca21} addressed the multipole response of weakly bound nuclei, shedding light on the origin of this low-lying component and clearing the path for experimental and theoretical developments up to date. They showed that changes in the binding energy for weakly bound systems not only shift the maxima in the low-lying multipole response, but have also important consequences for the total strength, a fact reflected by energy-weighted sum rules \cite{2005Na09}. This was not associated to the excitation of resonant states, but rather to an optimal matching between non-resonant continuum states and the weakly bound orbitals. Moreover, their analysis concluded that this low-lying strength does not come at the expense of the usual response at higher energies, so the mechanism involves an increase of the total strength available. It is now widely accepted that a smaller binding energy in loosely bound nuclei produces larger concentrations of multipole strength just above the continuum threshold. This feature is illustrated in Fig.~\ref{fig:binding} and has actual implications for nuclear reactions with exotic beams. 

\begin{figure}
	\centering
	\resizebox{0.9\columnwidth}{!}{%
		\includegraphics{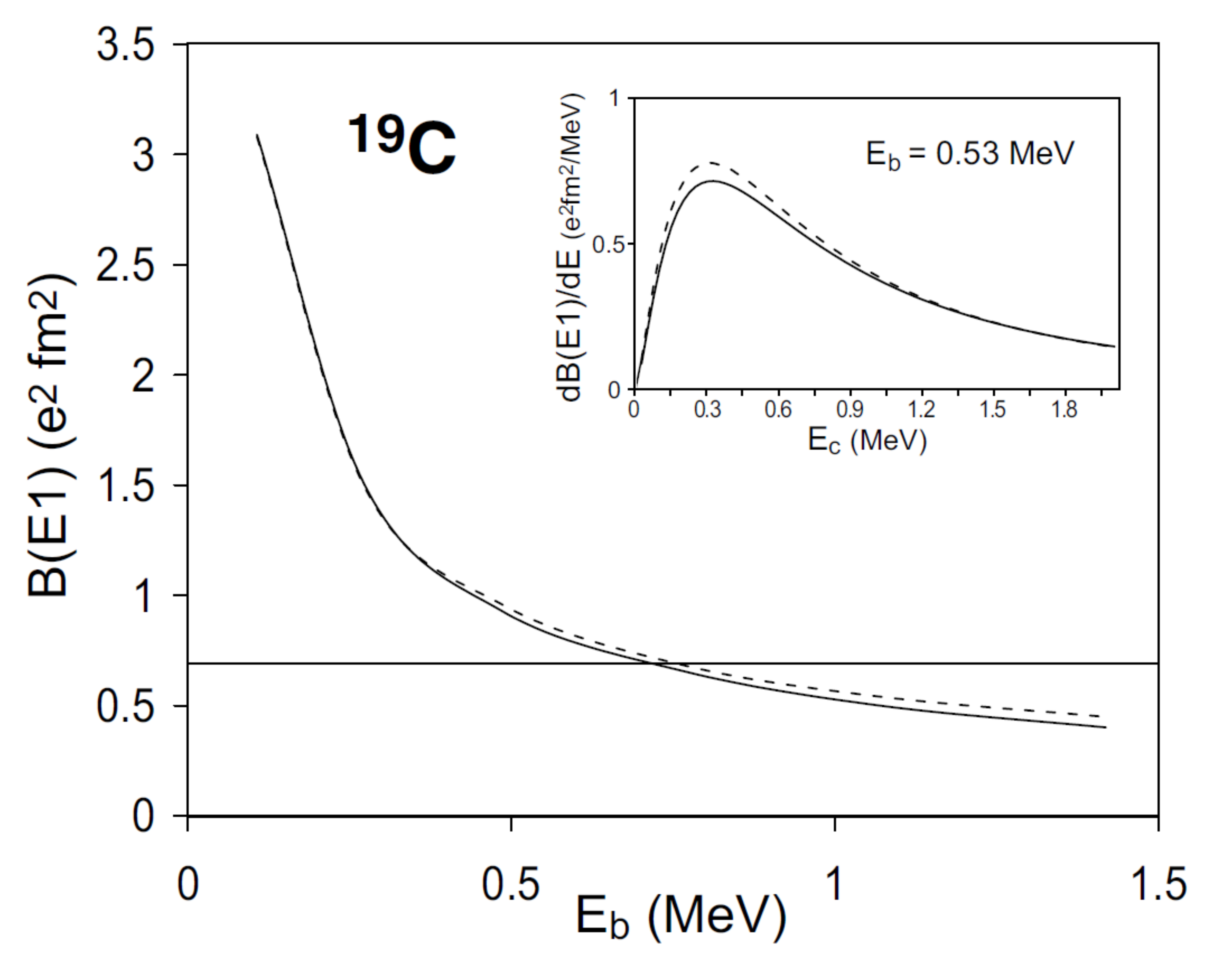}
	}
	\caption{Total $E1$ strength in $^{19}$C for transitions from a $2s$ bound orbital to the continuum, as a function of $E_b$. Solid: ``exact'' calculation using wave-functions from a Woods-Saxon potential. Dashed: analytic formula Eq.~(\ref{eq:totalE1}). The horizontal line is the experimental $B(E1)$ value, corresponding here to $E_b\sim0.53$ MeV. The inset shows the $E1$ distribution at that precise energy, whose maximum is at $E_c=3/5E_b$ according to Eq.~(\ref{eq:E1dist}). Figure from Ref.~\cite{2005Na09} with the authors' permission.}
	\label{fig:analytic}       
\end{figure}

In connection with breakup cross sections for halo nuclei being generally dominated by the low-lying dipole strength, in Ref.~\cite{2005Na09} a simple analytic description of the low-lying $E1$ distribution and total strength was derived for the case of weakly bound systems comprising a compact core plus one valence nucleon. The model considered pure single-particle transitions from the halo weakly bound state to continuum states described as plane waves (for neutrons) or Coulomb functions (for protons). For example, the $E1$ distribution in the case of initial $s$-wave neutrons into $p$-wave continuum was given by
\begin{equation}
{\left.\frac{dB(E1)}{dE}\right|}_{(s\rightarrow p)} = \frac{3\hbar^2(Z_{\rm eff}e)^2}{\mu\pi^2} \frac{\sqrt{E_b}E_c^{3/2}}{(E_b+E_c)^4},
\label{eq:E1dist}
\end{equation}
and the total strength,
\begin{equation}
B(E1)=\frac{3\hbar^2(Z_{\rm eff}e)^2}{16\pi} \frac{1}{\mu E_b},
\label{eq:totalE1}
\end{equation}
where $Z_{\rm eff}=-Z/A$, $\mu$ is the reduced core + neutron mass, and $E_b$ and $E_c$ are the energies of the bound state and continuum states, respectively. These simple expressions imply that the total strength is indeed approximately proportional to the inverse of the binding energy of the halo neutron and that the maximum of the $B(E1)$ distribution shifts to lower energies for a smaller binding energy. A particular example in the case of $^{19}$C ($^{18}$C + $n$) is illustrated in Fig.~\ref{fig:analytic}. In the case of proton halos, Coulomb corrections indicated a more gradual change. Similar conclusions were drawn by other studies (e.g., \cite{2002FO13}).

These ideas, originally devised for weakly bound nuclei characterized by single nucleon halos, were naturally extended to systems composed by loosely bound clusters. In Refs.~\cite{2005Fo15,2008Ma57,2009Ma11}, for instance, a dicluster model was developed to describe the electromagnetic response leading to cluster-cluster dissociation of the $A=7$ isobars  $^7$Li and $^7$Be. They showed that, in the case of weakly bound systems where the breakup is dominated by dissociation into two clusters of similar size, the main contribution comes from the nuclear quadrupole mechanism, since the dipole excitation is much smaller. Few-body models considering three or more clusters explicitly have followed over the years. Vitturi and collaborators have also been involved in three-body models to study multipole excitations, e.g., the study of the electric response of the halo nucleus $^{6}$He~\cite{2016Si19} and, more recently, $^{22}$C~\cite{2019Si00}. Their results confirm, in accord with the vast literature on the subject, that the effect of the binding energy on the low-lying strength discussed in Ref.~\cite{1996Ca21} applies also for systems comprising a compact core and several valence particles. 

Also in the context of few-body models and clustering, the description of $^{12}$C as three $\alpha$ particles has gained renewed attention due to the success of algebraic approaches in predicting and describing experimental spectra~\cite{2014MA37}. Vitturi and collaborators have recently focused on this matter, in particular studying selection rules for electromagnetic transitions~\cite{2016St17} and their link to inelastic excitation and breakup \cite{arXiv:1901.07954}. Work along this line is ongoing and will be the subject of future research.

\section{Heavy-ion sub-barrier fusion reactions} \label{FUS}

\begin{figure}[tb]
	\begin{center}
\resizebox{1.\columnwidth}{!}{ \includegraphics{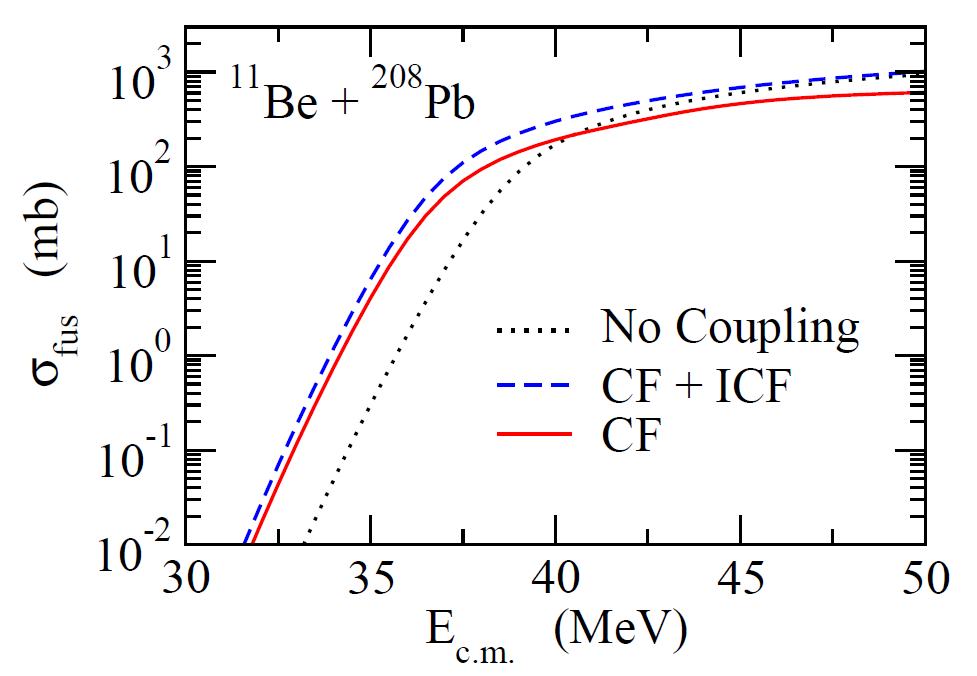}}
	\end{center}
	\caption{
		Fusion cross sections for the $^{11}$Be+$^{208}$Pb reaction as a function 
		of the incident energy in the center of mass frame. 
		The solid and the dashed lines are the complete fusion (CF) and the total 
		(i.e., the complete + the incomplete fusion) fusion cross sections, 
		respectively, obtained with the coupled-channels calculations which include 
		the breakup channels of the $^{11}$Be nucleus. 
		The dotted line shows the fusion cross sections in the absence of the 
		couplings. A redrawn of Fig. 2(a) in Ref. \protect\cite{2000HA14}.}
	\label{fig:bufusion}       
\end{figure}
In heavy-ion fusion reactions at energies 
around the Cou\-lomb barrier, it has been well recognized that 
inelastic excitations to 
low-lying collective states as well as several transfer processes 
lead to a large enhancement of fusion cross sections relative to 
the prediction of a simple potential 
model \cite{HT12,2015KU29,1997DA04,1995CH40,1989CA16,1987LO01}. 
A natural question then arises concerning 
how the breakup channel, which is important 
in weakly-bound nuclei, influences subbarrier fusion 
cross sections. One may naively think that the breakup channel 
reduces fusion cross sections, since a part of the incident flux 
is lost due to the breakup process. This idea was put forward by 
Hussein {\it et al.} \cite{1992HU03} as well as by 
Takigawa {\it et al.} \cite{1993TA04}. 
They computed fusion cross sections as, 
\begin{equation}
\sigma_{\rm fus}(E)=\frac{\pi}{k^2}\sum_l(2l+1)(1-P_{\rm bu}(E,l))P^{(0)}_{\rm fus}(E,l), 
\label{eq:bufusion}
\end{equation}
where $E$ is the incident energy in the center of mass frame and $k$ is 
the corresponding wave number. 
$P^{(0)}_{\rm fus}$ and $P_{\rm bu}$ are the penetrability of the Coulomb barrier 
in the absence of the breakup channel 
and the breakup probability for the partial wave $l$, respectively. 
That is, 
the authors of Refs. \cite{1992HU03,1993TA04} 
took into account, 
through the factor of $1-P_{\rm bu}$, 
the loss of flux due to the imaginary 
part of the dynamical polarization potential (DPP) for the breakup process.  
Naturally, the fusion cross sections are hindered in this formula 
as compared to the 
case with no breakup channel, that is, $P_{\rm bu}=0$. 

Andrea Vitturi, together with Carlos Dasso, pointed out that this naive picture 
does not hold \cite{1994DA13}. To this end, they used a simple two-level 
model, in which the breakup channel is represented as a 
single effective channel (see also Ref. \cite{2014KU03}), 
and estimated the probability for the 
complete fusion as the penetrability for the entrance channel. 
The resultant cross sections clearly showed an enhancement of fusion cross 
sections, rather than a hindrance, as compared to the no-coupling 
calculation. 
An interpretation for this result is that the coupling to the breakup 
channel dynamically lowers the Coulomb barrier, increasing the penetrability. 
That is, the DPP due to the breakup channel 
has both the real and the imaginary parts, and thus the 
probability $P^{(0)}_{\rm fus}(E,l)$ in Eq. (\ref{eq:bufusion}) has to be 
modified to $P_{\rm fus}(E,l)$ 
by taking into account the real part of the DPP. 

Subsequently, this calculation was further extended to a more 
realistic treatment for the breakup, done in collaboration of Andrea Vitturi with 
Kouichi Hagino, Carlos Dasso, and Silvia Lenzi \cite{2000HA14}. 
To this end, they discretized the continuum states in bins of energy 
and associated with each bin the coupling form factor corresponding 
to its central energy. The coupling form factors, both for the 
nuclear and the Coulomb parts, were evaluated 
microscopically using the core+neutron model, which took into account 
the weak bound nature of a projectile nucleus. 
The continuum-continuum couplings were not taken into account. 
As in Ref. \cite{1994DA13}, the complete (incomplete) 
fusion was identified as those 
flux penetrating the Coulomb barrier in the entrance channel 
(the breakup channels). 
Notice that this calculation yields only the lower bound of the complete 
fusion, since both of the breakup fragments may be absorbed by the target 
nucleus when the breakup takes place inside the barrier, contributing to the 
complete fusion. 

The result for the $^{11}$Be+$^{208}$Pb reaction is shown in 
Fig. \ref{fig:bufusion}. 
The solid and the dashed lines show the results for the complete 
and the total fusion cross sections, respectively, while the dotted 
line shows the result in the no-coupling limit. 
One can clearly see that the cross sections for complete fusion 
are enhanced 
at energies below the barrier while they are hindered at above barrier 
energies, as compared to the no-coupling case shown by the dotted line. 
It is evident that these are due to the real and the imaginary parts 
of DPP, respectively.
The continuum-continuum couplings would reduce the degree of enhancement 
of complete fusion cross sections at subbarrier energies \cite{2002DI02}, 
but one could still 
expect some enhancement at energies well below the Coulomb barrier. 

In this connection, Andrea Vitturi has made important contributions 
to the field as shown in two short papers with Carlos Dasso and 
Manuel Lozano published in Phys. Rev. A \cite{DLV91,DLV92}. 
In the first paper \cite{DLV91}, they discussed the role of a closed channel 
in multi-channel penetrabilities. 
One may have thought that a closed channel is kinematically forbidden and 
thus does not affect physical processes. 
Of course, this is not the case, since virtual excitations to a closed 
channel is still possible, which may affect significantly e.g., a barrier 
penetrability. Using a simple two-level model, they explicitly demonstrated 
that the penetrability is enhanced by the channel coupling effects, even when 
the second channel is closed. This result can actually 
be interpreted in terms of 
barrier distribution \cite{HT12} such that the coupling leads to a splitting 
of the barrier, yielding a lower barrier than the uncoupled one, 
irrespective to the value of the incident energy. 
In the second paper \cite{DLV92}, they discussed the applicability of the 
adiabatic approximation, which is valid when the energy of an excited 
channel is large. In this approximation, 
the channel coupling effect leads to a renormalization of 
potential \cite{HT12,1994TA07}. They pointed out that the applicability 
of the adiabatic approximation depends not only on the energy of the 
excited state but also on the coupling strength. 
That is, for a finite excitation energy, the adiabatic approximation may 
break down if the coupling strength is large. 
The same conclusion can be obtained also by evaluating a correction factor 
to the adiabatic approximation \cite{1994TA07,1995TA01}. 
We point out that these conclusions of the two papers of Andrea Vitturi
are considerably important when one discusses 
astrophysical fusion reactions.

\subsection{Fusion at astrophysical energies in light clusterized systems} \label{Astro}
\begin{figure}[tb]
	\begin{center}
		\resizebox{1.\columnwidth}{!}{ \includegraphics{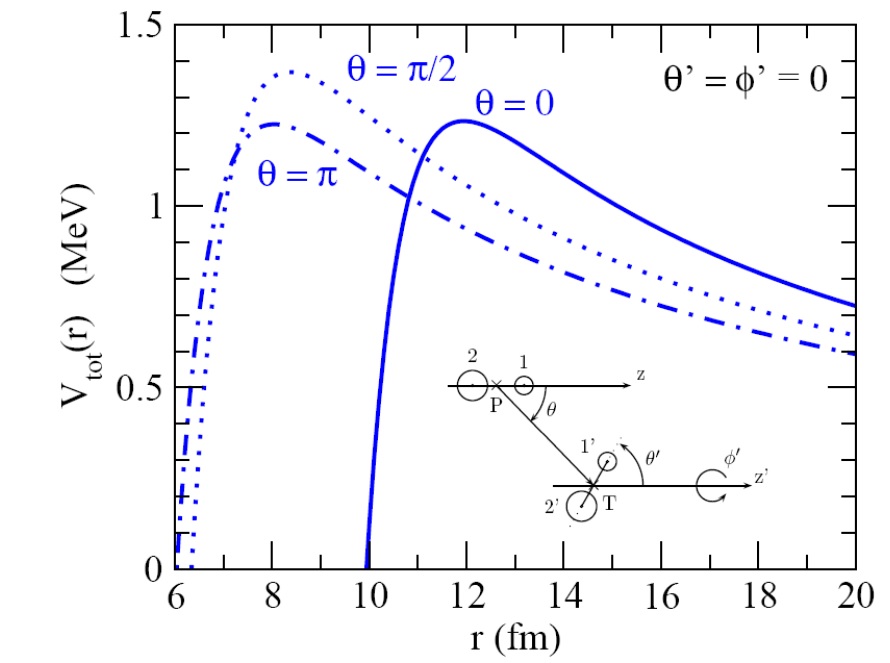}}
	\end{center}
	\caption{ Potential barriers for $^6$Li + $^6$Li fusion, where the cluster structure ($\alpha$+$d$) as been taken into consideration (see the inset for explanation of the coordinates).	}	
	\label{fig:screen}       
\end{figure}
A recent work that was initially proposed by C.A.Bertulani and C.Spitaleri and finished with the contribution of the Padova group \cite{2016Sp00} treats the problem of fusion at astrophysical energies for light clusterized nuclei. At the temperatures at which fusion occurs in stars, the fusion rate is enhanced by the presence of a gas of electrons in the stellar plasma, that shows significant deviations with respect to the fusion cross-section measured in laboratory experiments, leading to the necessity to include an effective screening potential. A reason for the so-called electron screening problem \cite{1987Ass,Chapter_Lang}, that has been a long-standing issue in this field, has been traditionally  attributed to atomic properties, but a comprehensive theory was not yet available.
In Ref.  \cite{2016Sp00}, the reason for the anomalous values of the effective screening potential was attributed to nuclear clustering phenomena, that might severely affect the fusion cross-sections in reactions involving light nuclei. In a WKB approach, the tunneling probability can be calculated as 
\be 
P=e^{-2G}
\ee
where the Gamow factor depends on the potential between target and projectile. Giving a clusterized substructure to one of the nuclei involved (or to both) leads to potential of a composite nature that depend not only on the distance but also on the relative orientation of the clusters. This implies that the Gamow factor 
\be
G(E,\theta,\theta',\phi')=\frac{\sqrt{2m}}{\hbar} \int_a^b \sqrt{V_{tot}(r,\theta,\theta',\phi')-E} ~dr 
\ee
has a dependence on the angles that generates strong changes in the probability as depicted in Fig. \ref{fig:screen}, where the coordinate system and potential are given in the case of a dicluster nucleus impinging on another dicluster nucleus ($^6$Li on $^6$Li in this case). Here $a$ and $b$ are inversion points of classical motion.

The Coulomb barrier is suppressed and shifted by the presence of clusters and polarization and this fact must be taken into account when modeling the interaction between light nuclei in the fusion process as it certainly has profound implications on the barrier penetrability and fusion rates.  

%
%
%
%

\section{Authors contributions}
All the authors were involved in the preparation of the manuscript.
All the authors have read and approved the final manuscript.
We thank M. B\"oy\"ukata for preparation of Fig. 4.

\end{document}